\documentclass[twocolumn,prc,showpacs,superscriptaddress,floatfix]{revtex4}
\usepackage{graphicx}

\begin{document}

\title{
Systematic study of three-nucleon force effects\\ 
in the cross section of the deuteron-proton breakup at 130 MeV}

\author{St.~Kistryn}
\email[Electronic mail: ]{skistryn@if.uj.edu.pl}
\affiliation{Institute of Physics, Jagiellonian University,
             PL-30059 Krak\'ow, Poland}
\author{E.~Stephan}
\affiliation{Institute of Physics, University of Silesia, 
             PL-40007 Katowice, Poland}
\author{A.~Biegun}
\affiliation{Institute of Physics, University of Silesia, 
             PL-40007 Katowice, Poland}
\author{K.~Bodek}
\affiliation{Institute of Physics, Jagiellonian University,
             PL-30059 Krak\'ow, Poland}
\author{A.~Deltuva}
\affiliation{Centro de Fisica Nuclear da Universidade de Lisboa,
             P-1649-003 Lisboa, Portugal}
\author{E.~Epelbaum}
\affiliation{Jefferson Laboratory, Theory Division, 
             Newport News, VA 23606, USA}
\author{K.~Ermisch}
\affiliation{Kernfysisch Versneller Instituut, 
             NL-9747 AA Groningen, The Netherlands}
\author{W.~Gl\"ockle}
\affiliation{Institut f\"ur Theoretische Physik II,
             Ruhr Universit\"at Bochum, D-44780 Bochum, Germany}
\author{J.~Golak}
\affiliation{Institute of Physics, Jagiellonian University,
             PL-30059 Krak\'ow, Poland}
\author{N.~Kalantar-Nayestanaki}
\affiliation{Kernfysisch Versneller Instituut, 
             NL-9747 AA Groningen, The Netherlands}
\author{H.~Kamada}
\affiliation{Department of Physics, Kyushu Institute of Technology, 
             1-1 Sensucho, Tobata, Kitakyushu 804-8550, Japan}
\author{M.~Ki\v{s}}
\affiliation{Kernfysisch Versneller Instituut, 
             NL-9747 AA Groningen, The Netherlands}
\author{B.~K{\l}os}
\affiliation{Institute of Physics, University of Silesia, 
             PL-40007 Katowice, Poland}
\author{A.~Kozela} 
\affiliation{Institute of Nuclear Physics Polish Academy of Science, 
             PL-31342 Krak\'ow, Poland}
\author{J.~Kuro\'s-\.Zo{\l}nierczuk}
\altaffiliation{Present address: Nuclear Theory Center, 
             Indiana University, Bloomington, IN 47405 USA}
\affiliation{Institute of Physics, Jagiellonian University,
             PL-30059 Krak\'ow, Poland}
\author{M.~Mahjour-Shafiei}
\altaffiliation{Present address: Department of Physics, 
                University of Tehran, Tehran 1439955961, Iran}
\affiliation{Kernfysisch Versneller Instituut, 
             NL-9747 AA Groningen, The Netherlands}
\author{U.-G.~Mei{\ss}ner}
\affiliation{Universit\"at Bonn, Helmholtz-Institut f\"ur
             Strahlen- und Kernphysik (Theorie), D-53115 Bonn, Germany}
\affiliation{Forschungszentrum J\"ulich, Institut f\"ur Kernphysik
             (Theorie), D-52425 J\"ulich, Germany}
\author{A.~Micherdzi\'nska}
\altaffiliation{Present address:
             Indiana University, IUCF, Bloomington, IN 47405 USA}
\affiliation{Institute of Physics, University of Silesia, 
             PL-40007 Katowice, Poland}
\author{A.~Nogga}
\affiliation{Forschungszentrum J\"ulich, Institut f\"ur Kernphysik
             (Theorie), D-52425 J\"ulich, Germany}
\author{P.~U.~Sauer}
\affiliation{Institut f\"ur Theoretische Physik, Universit\"at
             Hannover, D-30167 Hannover, Germany} 
\author{R.~Skibi\'nski}
\affiliation{Institute of Physics, Jagiellonian University,
             PL-30059 Krak\'ow, Poland}
\author{R.~Sworst}
\affiliation{Institute of Physics, Jagiellonian University,
             PL-30059 Krak\'ow, Poland}
\author{H.~Wita{\l}a}
\affiliation{Institute of Physics, Jagiellonian University,
             PL-30059 Krak\'ow, Poland}
\author{J.~Zejma}
\affiliation{Institute of Physics, Jagiellonian University,
             PL-30059 Krak\'ow, Poland}
\author{W.~Zipper}
\affiliation{Institute of Physics, University of Silesia, 
             PL-40007 Katowice, Poland}

\date{\today}

\begin{abstract}
High precision cross-section data of the deuteron-proton breakup
reaction at 130 MeV are presented for 72 kinematically complete
configurations.  The data cover a large region of the available 
phase space, divided into a systematic grid of kinematical variables.  
They are compared with theoretical predictions, in which the full 
dynamics of the three-nucleon (3$N$) system is obtained in three 
different ways: realistic nucleon-nucleon ($NN$) potentials are 
combined with model 3$N$ forces (3NF's) or with an effective 3NF 
resulting from explicit treatment of the $\Delta$-isobar excitation. 
Alternatively, the chiral perturbation theory approach is used at 
the next-to-next-to-leading order with all relevant $NN$ and 3$N$ 
contributions taken into account.  The generated dynamics is then 
applied to calculate cross-section values by rigorous solution of 
the 3$N$ Faddeev equations.  The comparison of the calculated cross
sections with the experimental data shows a clear preference for the 
predictions in which the 3NF's are included. 
The majority of the experimental data points is well reproduced by the
theoretical predictions.  The remaining discrepancies are investigated
by inspecting cross sections integrated over certain kinematical 
variables.  The procedure of global comparisons leads to establishing 
regularities in disagreements between the experimental data and the 
theoretically predicted values of the cross sections.
They indicate deficiencies still present in the assumed models 
of the 3$N$ system dynamics.
\end{abstract}


\maketitle

\section{\label{secI}Introduction}

The dynamics of the three-nucleon (3$N$) system can be very accurately
studied by means of the nucleon-deuteron breakup reaction.  Its final 
state, constrained by only general conservation laws, provides a
rich source of information to test the nuclear Hamiltonian.  
It is of particular importance for components of the models which account
for subtle effects, like three-nucleon force (3NF) contributions to 
the potential energy of the 3$N$ system.  
Precise predictions for observables in the 3$N$ system can be 
obtained via exact solutions of the 3$N$ Faddeev equations for 
any nucleon-nucleon ($NN$) interaction, even with the inclusion 
of a 3NF model~\cite{Glo96a}.  
To investigate details of the dynamics of the 3$N$ system, in addition 
to elastic $Nd$ scattering data, reliable deuteron breakup data 
sets, covering large regions of the available phase space, are needed.  
Unfortunately, it still remains difficult to perform 
such measurements at the required level of precision.  In our previous
paper~\cite{Kis03a} we have started to report results of a project 
dedicated exactly towards such an aim.  Here we continue with the 
presentation of a systematic set of breakup cross-section values 
and we compare them with theoretical predictions based on various 
dynamical assumptions.

Properties of few-nucleon systems at not-too-high energies are 
determined mainly by pairwise nucleon-nucleon interactions.  
Models of $NN$ forces describe the long range interaction part 
according to the meson-exchange picture, while the short range 
is based on phenomenology, and adjusted by fitting a 
certain number of parameters to the $NN$ scattering data.  The present
generation of $NN$ potentials reaches an unprecedented accuracy in
describing the $pp$ and $np$ observables below 350 MeV, expressed
by a $\chi^2$ per degree of freedom very close to 1.  In few-nucleon 
studies the most widely used so-called ``realistic'' $NN$ potentials 
are Argonne $\upsilon_{18}$ (AV18)~\cite{AV18}, charge dependent 
(CD)~Bonn~\cite{CDBonn1,CDBonn2}, Nijmegen I and II (Nijm~I, 
Nijm~II)~\cite{NijmIII}.  Their full equivalence with  
phase shift analysis~\cite{NPSA} guarantees that all two-body 
aspects of the interaction are taken into account when these $NN$ force 
models are used in microscopic calculations of few- and many-nucleon
systems.

At the more fundamental level of Quantum Chromodynamics (QCD), the 
strong force between the nucleons is understood as residual color force. 
A direct description of few-nucleon systems at low energy from first 
principles would require the solution of QCD in the non-perturbative 
regime which is not possible at present (except on the lattice). 
On the other hand, the low-energy dynamics of QCD can be studied in 
the chiral effective field theory (EFT) framework. This is a systematic 
approach which incorporates the spontaneously broken approximate chiral 
symmetry of QCD and is based on the most general effective Lagrangian 
for Goldstone bosons (pions in the two-flavor sector of the up and down 
quarks) and matter fields (nucleons, $\Delta$--resonances, $\ldots$). 
In the pion and single-baryon sectors, $S$-matrix elements can be 
calculated in chiral perturbation theory (ChPT) via an expansion 
in terms of $(Q/\Lambda_{\chi})^{\nu}$ -- in powers $\nu$ of a 
low-momentum scale $Q$, associated with small generic external 
momenta and with the pion (light-quark) mass.  Here, small means with 
respect to the scale $\Lambda_{\chi}$, corresponding to the chiral 
symmetry breaking scale of the order of 1 GeV. 
Motivated by successful applications 
of ChPT in the $\pi \pi$ and $\pi N$ sectors, Weinberg proposed to extend 
the formalism to systems with two and more nucleons, where non--perturbative 
calculations are necessary to deal with the shallow bound states (or large 
scattering lengths)~\cite{Wei79a,Wei90a}.  According to Weinberg, 
ChPT can be applied in that case not to the amplitude but to a kernel of the 
corresponding dynamical equation which may be viewed as an effective nuclear 
potential.  Few-nucleon $S$-matrix elements are generated non-perturbatively 
by iterating the potential in the dynamical equation.  The first application 
of this approach in the 2N sector was performed in~\cite{Ord92a,Ord94a}. 
At present, the 2N system has been studied up to 
next-to-next-to-next-to-leading order (N$^3$LO) in the chiral 
expansion~\cite{Ent03a,Epe05a}, while the three- and more numerous 
nucleon systems have so far been analyzed up to next-to-next-to-leading 
order (NNLO)~\cite{Epe02a,Ent02a,Epe04a}.  It should be stressed that 
this approach offers an unique possibility to estimate uncertainties
of the theoretically predicted physical quantities.

High-quality models of the $NN$ potentials, when applied to calculate
observables in the 3$N$ system, revealed discrepancies between the
pure pairwise dynamics and the experimental results.  The most
promising and widely investigated explanation is the presence of
three-nucleon interactions.  The realistic potentials are
therefore supplemented by 3NF models, usually refined versions
of the Fujita-Miyazawa force~\cite{FM3NF}, in which one of the 
nucleons is excited into an intermediate $\Delta$ via a 2$\pi$-exchange
between both remaining nucleons.  The most popular version of such an 
interaction is the Urbana IX~\cite{UIX3NF} force.  
The Tucson-Melbourne (TM)~\cite{TM993NF} 3NF extends this picture by
allowing for additional processes contributing to the pion rescattering
at the intermediate nucleon.
An alternative mechanism of generating a 3NF is based on the so-called 
explicit $\Delta$-isobar excitation~\cite{Nem98a,Chm03a,Del03a,Del03b}. 
Calculations are performed in a coupled-channel approach and the 
effective 3NF is generated (together with other $\Delta$-isobar effects) 
due to the explicit treatment of the degrees of freedom of a single 
$\Delta$.  Finally, within the ChPT framework both 2$N$ and 3$N$ forces 
(as well as nuclear currents) are derived from the same effective chiral 
Lagrangian and are thus fully consistent with each other.  This leads to 
a consistent model of the $NN$ and 3$N$ interactions, which also strongly 
constrains the parameters of the 3NF.  As stated earlier, presently the 
results in the 3$N$ system are only available at NNLO.  The analysis at 
N$^3$LO requires sophisticated analytical and numerical calculations. 
This work is in progress.

The role of 3NF effects has been recognized already when
studying the bound states of three nucleons.  No realistic 
potential approach can reproduce the binding energies of $^3$He 
and $^3$H when the calculations are based on $NN$ forces
only~\cite{Nog00a}.  When 3NF contributions are taken into account, 
the $^3$H and $^3$He binding energies can be described accurately
(by construction, because parameters of the 3NF are usually fitted
to match the triton binding) -- see e.g.~\cite{Nog03a}.  These combined
models of $NN$ and 3$N$ forces also describe the $^4$He binding 
energy, indicating that 4NF are presumably small~\cite{Nog02a}.
For the description of the level schemes of p-shell nuclei, the most 
simple 3NF's show failures which motivated more sophisticated 3NF 
models leading to encouraging agreement between theory and 
experiment~\cite{Pie01a}.  Here, we will restrict ourselves only 
to the models mentioned above.  In the isospin $T$ = 1/2 state, they 
are expected to be very similar to the extended versions used 
in~\cite{Pie01a}.
An analogous conclusion is obtained within the ChPT framework -- inclusion
of 3NF graphs leads to an improved description of few-nucleon bound 
states~\cite{Epe02a}.  Further evidence of relevant
consequences originating from introducing additional dynamics into 
the 3$N$ system comes from the coupled-channel approach -- the binding 
energies of $^3$He and $^3$H are much closer to the experimental values
when the $\Delta$-isobar contributions are included and the difference 
of the two bindings is well matched~\cite{Del03b}.

Presently, the richest evidence for the importance of 3NF effects is 
deduced from the elastic nucleon-deuteron scattering observables.  The 
picture emerging from the comparisons of various data with theory is, 
however, rather ambiguous. In several cases where the $NN$ forces alone 
fail to reproduce the observables the inclusion of 3NF's leads to significant
improvements~\cite{Wit98a,Nem98b,Sak00a,Erm01a,Cad01a,Hat02a,Erm03a,Sek04a,Mer04a} (for earlier references c.f.~\cite{Glo96a}).
Alas, in several cases discrepancies between the experimental data
and theoretical predictions remain, even if the presently available 
full 3$N$ dynamics is taken into account.  This statement is 
especially true for various polarization 
observables~\cite{Sak00a,Erm01a,Hat02a,Sek04a}, but holds also for 
certain cross-section angular distributions (see e.g.~\cite{Erm03a}).  
Those failures, confirmed by different calculational approaches, 
indicate that the 3NF models are still missing some relevant ingredients, 
while for the ChPT framework they might suggest the necessity for 
including higher order (at least N$^3$LO) terms for the 3$N$ system.  

Since the theoretical models clearly need more constraints from the
experimental data, it is natural to extend the investigations
of the 3$N$ system to the nucleon-deuteron ($Nd$) breakup reaction.  
The continuum of the final states, which has to be simultaneously 
described in its full richness by the assumed dynamical model of $NN$ 
and 3$N$ interactions, should provide a lot of information to pin down 
the details of the theoretical models.  Unfortunately, this field 
has hardly been explored experimentally and only at lower energies, 
below 30 MeV nucleon energy (see \cite{Glo96a} and \cite{Kur02b} for 
references; the most recent results can be found in~\cite{Set05a}
and~\cite{Duw05a}, and in~\cite{Mey04a} at much higher incident energy).
In the region of intermediate energies (30 MeV -- 100 MeV) only at
65 MeV several isolated kinematical configurations have been 
investigated with respect to cross sections and analyzing 
powers~\cite{All94a,All96a,Zej97a,Bod01a}.  Comparison of those data 
with the theoretical predictions obtained within the 
approaches~\cite{Kur02b,Epe02a,Del03a,Del03b} discussed above 
shows again a mixed picture: sometimes the agreement is improved by 
including 3NF's, in some cases the 3NF effect is negligible and there 
are cases in which inclusion of 3NF's moves the prediction away from 
the data.  Since the thorough theoretical study~\cite{Kur02a} of the 
full phase space of the breakup reaction shows that significant effects
can be expected, there is a strong need for data which precisely and 
systematically scan large ranges of the final state kinematical
variables.  

Therefore, we have performed a $^1$H($\vec{\textrm{d}}$,pp)n 
breakup experiment using a beam of $130$ MeV polarized deuterons 
(equivalent to 65 MeV incident nucleon energy).  The experiment
has been performed at KVI in Groningen, employing a detector setup
covering a large fraction of the full breakup phase space. High 
precision cross sections together with vector and tensor analyzing 
powers have been measured in kinematically complete configurations by
registering energies and angles of the two outgoing coincident protons.
We have already reported~\cite{Kis03a} a comparison of the first set 
of the breakup cross sections with realistic $NN$ forces and 3NF model
predictions, finding unambiguously significant effects of the 
three-nucleon interaction.  In this paper we present an extended breakup
cross-section data set for 72 kinematical configurations (corresponding
to a total of about 1200 data points).  
Since we have introduced certain improvements in the data analysis 
procedure, this set partially overlaps with the previous data.  A second 
reason for such an overlap is to provide now a fully systematic coverage 
of the phase space, presenting the data on a grid of kinematical variables
(two proton polar angles, their relative azimuthal angle and the arclength 
variable).  We compare our experimental results to theoretical predictions 
based on various approaches.  First, we use realistic $NN$ potentials 
combined with phenomenological 3$N$ interactions.  Then, we base the
predictions on a coupled-channel potential with the explicit single 
$\Delta$-isobar degrees of freedom.  Finally, we use also the results
of the calculations within the ChPT framework at NNLO, with complete 
2$N$ and 3$N$ dynamics.  The comparison is supplemented by first global 
searches of possible regularities in differences between the data and 
theory, determined by inspecting the cross sections summed over certain 
kinematical variables.     
 
There are a few issues which need to be discussed in order to clarify
the details of interpreting the experimental results.  First of all,
we already discussed that the 3$N$ interaction is still not completely
understood.  Recent studies of strongly non-local interactions 
(see~\cite{Dol03a,Dol04a}) aim at total removal of the 3NF's.  Indeed, 
the non-locality is closely related to the 3NF's~\cite{Pol90a} and,
in principle, this can result in an ambiguous separation of 3NF effects
and off-shell effects.  Here we will not discuss this issue further.  
We only note that our predictions are based on several $NN$ interactions,
some of which are local, some are non-local.  Nevertheless, they all 
provide very similar predictions, alone and when combined with
model 3NF's.  We also note that the chiral interactions we use are 
evidently non-local.  It should also be mentioned
that all the applied formalisms miss two features which are inherently
present in the experimental data.  The first difference is the
Coulomb interaction: the experiment is performed in the deuteron-proton
system while all calculations neglect any long-range forces like the
Coulomb interaction.  It can be
argued that the influence of the Coulomb interaction at our energy
is (if any) very small.  Calculations for the elastic scattering
cross section at 65 MeV~\cite{Kie04a,Del05b} indicate an essentially 
negligible difference for $nd$ and $pd$ predictions, even in the 
cross-section minimum, the most sensitive region to study the 3NF 
effects.  The simultaneous treatment of the Coulomb and nuclear forces
in the Faddeev framework is progressing~\cite{Del05a},
but predictions for our breakup data are not available yet.  
The first information suggest, however, that in 
contrast to the elastic scattering case, the Coulomb effects can 
significantly influence the breakup cross sections in certain 
kinematical configurations. 
Secondly, all the theoretical approaches are using a nonrelativistic
framework and nonrelativistic kinematics.  Here again we expect the
effects induced by relativity to be almost negligible.  For cross
sections of the breakup reaction in selected configurations it has
been shown~\cite{Fac03a} using relativistic kinematics that the 
differences between both treatments are minimal at nucleon energies 
below 100 MeV.  The remaining problem of arclength differences does 
not introduce any noticeable effects either -- we adopt a projection 
procedure~\cite{All94a}, transforming the theoretical predictions onto 
the relativistic kinematics.  Similar conclusions were also reached 
in~\cite{Chm03a}.  Ultimately rigorous comparison will be possible 
only after a full relativistic dynamics (boosted potentials) is 
implemented for the Faddeev formalism, similarly to the first
calculations for the 3$N$ bound states~\cite{Kam02a} and to 
the pioneering $Nd$ elastic scattering study~\cite{Wit05a}.

The paper is organized as follows. In Section~\ref{secII} we
recall some details of the experiment and of the data analysis,
emphasizing the refinement introduced since the previous 
report~\cite{Kis03a}. We briefly present in Section~\ref{secIII} 
the theoretical formalism underlying the calculations based on
solving the Faddeev equations with the realistic potentials, with 
the effective potential obtained in the ChPT framework and the 
coupled-channel approach with the explicit $\Delta$-isobar
excitation treatment.  Our high precision breakup cross-section 
data are presented and compared to theoretical predictions in 
Section~\ref{secIV}.  We conclude and summarize in Section~\ref{secV}.

\section{\label{secII}Experiment and Data Analysis}

\subsection{\label{secIIA}Setup and measurement procedure}

The experiment was performed at the Kernfysisch Versneller Instituut 
(KVI), Groningen, The Netherlands.  Only the main features of the
experimental procedure are briefly summarized in the following,
the detailed description can be found elsewhere~\cite{Kis03a,MichPhD}.

\begin{figure}[t]
\includegraphics[width=86mm]{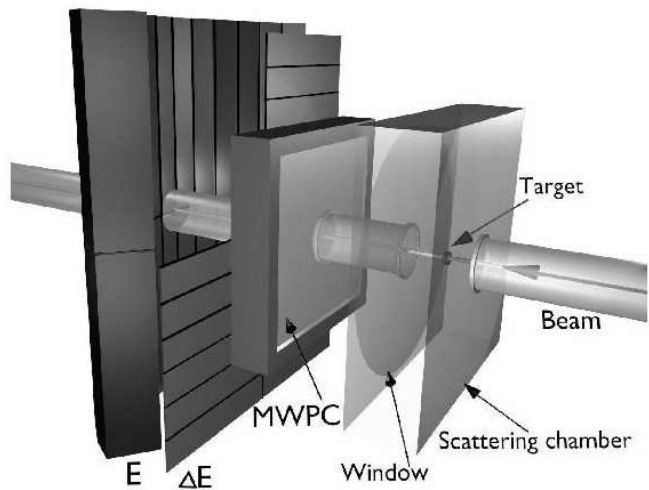}
\caption{\label{fig_setup}
Schematic view of the detection system, presenting the relative positions 
of the MWPC, the two layers of the scintillation detectors and the scattering 
chamber containing the target.  The central beam line transporting the 
primary beam to the target and further downstream to a distant Faraday
cup is also indicated.  For the sake of better view on the $E$-detector
wall, 6 $\Delta E$ detectors (one sector) are removed from the figure.}
\end{figure}

The beam of vector and tensor polarized deuterons was focused to a spot 
of approximately 2 mm diameter on a liquid hydrogen target of few mm 
thickness. The SALAD (small angle large acceptance detector)~\cite{Kal00a} 
detection system consisted of a three-plane multiwire proportional chamber 
(MWPC) and two layers of scintillator hodoscope (cf.~Fig.~\ref{fig_setup}). 
The MWPC was used for precise reconstruction of the charged-particle 
emission angles.  To resolve reconstruction ambiguities for multihit 
events, the MWPC consisted of three active anode planes with wires spanned 
horizontally ($x$), vertically ($y$) and diagonally ($u$).  The almost 
point-like reaction region, as compared to the target-MWPC distance, 
allowed for reconstruction of the polar and azimuthal particle emission 
angles with the overall accuracy of 0.6$^{\circ}$.

The plastic scintillator hodoscope covered the range of polar angles 
between 10$^{\circ}$ and 35$^{\circ}$, and the full range of azimuthal
angles. It consisted of 24 transmission detectors (horizontal $\Delta E$ 
strips) and 24 stopping detectors (vertical $E$ slabs), forming together 
a two-dimensional array of 140 elements, with an area of about 
60$\times$60~mm$^{2}$ each.  The system possessed mirror symmetries
with respect to the horizontal and vertical planes, i.e.~it could be
viewed as composed of four similar sectors, each consisting of 6 slabs 
and 6 strips.  Strips belonging to one sector formed telescopes with 
slabs of the same sector, while they had no overlap with slabs in other 
sectors.  This physical grouping of detectors had a reflection in
the trigger logic, based on combination of hit multiplicities within 
the sectors.  Apart from trigger definition, information from the
telescope array was used for particle identification and for 
determination of their energies.

The events of interest can be roughly divided into three classes.
First we distinguish single events, for which only one 
$\Delta E$--$E$ telescope of the scintillation array has registered 
signals in a proper time window.  The other two types are
coincident events with two particles detected in two different
telescopes.  Among them a distinction was made between coincidences
of elements belonging to the diagonal sectors (candidates for both, 
elastic scattering and breakup events) and elements of the adjacent 
sectors or belonging to the same sector (only breakup events).  
These three kinds of triggers were separately downscaled, enhancing 
the coincidence rates, to a level acceptable for the data acquisition 
system.  Fine classification of events has been done off-line by 
incorporating the MWPC information -- for an extensive description 
see~\cite{MichPhD}.

For each registered event the information from the readout system 
comprised data from the scintillator hodoscope and from the MWPC.
The hodoscope data included times measured with respect to the
cyclotron reference (rf) signal and pulse heights for all
active detectors (strips and slabs).  The MWPC information was coded into
the numbers of the hit wires (more precisely -- centers and widths of 
the adjacent groups, i.e. clusters, of wires).  In addition, several 
auxiliary pieces of information were stored with each event: the beam 
polarization state, trigger pattern at various electronic stages, etc.  
Scalers, trigger rates, integrated beam current, pulse generator
signals for dead-time monitoring, etc., were read out every 1~s.

\subsection{\label{secIIB}Data analysis}

All basic steps of the data analysis procedure, like event selection, 
energy calibration, determination of detection efficiencies and cross 
section normalization, have been thoroughly described in~\cite{Kis03a}.
The description below recalls the main features with an emphasis put on
the introduced improvements and on additional studies performed with 
the aim to reduce experimental uncertainties or to control their 
magnitude with enhanced accuracy. 

\subsubsection{\label{secIIB1}Selection of events and background subtraction}

The first step of the analysis was an adequate selection of the events of 
interest, i.e.~coincident proton - proton pairs from the breakup process or, 
necessary for cross section normalization, deuteron - proton coincidences 
originating from the elastic scattering.  To guarantee that only the 
products of the reactions initiated within a single beam burst were selected, 
a 20-ns-wide time window was imposed on the time spectra. Particle 
identification, based on the $\Delta E$--$E$ technique, proved to be 
very reliable, providing very good separation between protons and 
deuterons in the whole energy range.

Energy calibration was performed on the basis of data collected in special 
calibration runs with energy degraders of precisely known thicknesses.
The positions of the peaks corresponding to protons from elastic 
scattering which traversed the degraders were compared with the results 
of simulations taking into account all energy losses of protons on their 
paths from the reaction point to the detectors.  In this way the relation 
between ADC conversion (pulse height) and the energy deposited in the $E$ 
counter was found. The relation between the deposited energy and the proton 
energy at the moment of the reaction was obtained by analogous simulation 
of the proton energy losses.  With all these provisions for each breakup 
event the initial energies of both protons ($E_{1}$ and $E_{2}$) were 
determined.  
 
\begin{figure}
\includegraphics[width=86mm]{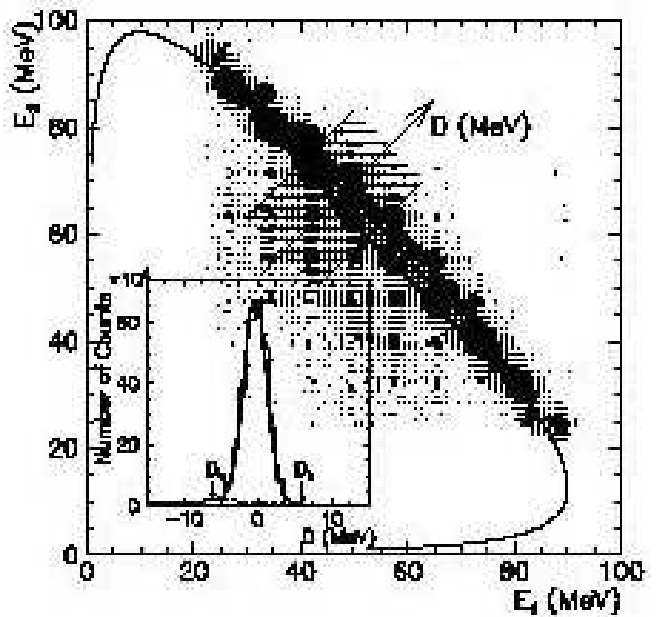}
\caption{\label{fig_e1e2}
$E_1$ versus $E_2$ coincidence spectrum of the two protons registered 
at $\theta_{1}$ = 20$^{\circ}\pm1^{\circ}$, 
$\theta_2$ = 15$^{\circ}\pm1^{\circ}$ 
and $\phi_{12}$ = 100$^{\circ}\pm5^{\circ}$.
The solid line shows a 3-body kinematical curve calculated for the 
central values of the experimental angular ranges.
Inset: determination of the accidental coincidences.
The spectrum was obtained by choosing one slice along the kinematical
curve in the $E_1$ vs.\ $E_2$ spectrum (marked area in the main figure) 
and projecting the events onto the $D$ axis.
Solid line represents the sum of a linear background function (shown
with the dashed line) and a Gaussian distribution, fitted in the range 
of $D$ between $D_{\textrm{a}}$ and $D_{\textrm{b}}$, corresponding 
to distances of $-3\sigma$ and $+3\sigma$ from the peak position.}
\end{figure}

The coincidence (kinematic) spectra $E_{1}$ vs.\ $E_{2}$ were built for 
each analyzed configuration, defined by polar angles $\theta_1$, $\theta_2$
and relative azimuthal angle $\phi_{12}$ of the two emitted protons.
The integration limits of $\Delta\theta_1 = \Delta\theta_2 = 2^{\circ}$ 
and $\Delta\phi_{12} = 10^{\circ}$ were used in all experimental
integrations leading to the cross-section results, as well as in
the studies concerning the performance of the detection system. 
The energies $E_{1}$ and $E_{2}$ of each event were transformed into two 
new variables: $D$, denoting the distance of the $(E_{1},E_{2})$ point 
from the kinematical curve in the $E_{1}-E_{2}$ plane, and $S$ -- the value 
of the arclength along the kinematics.  Events in slices along the $S$
axis were projected on the central $D$ axis, as shown in 
Fig.~\ref{fig_e1e2}. In the resulting spectra (inset in 
Fig.~\ref{fig_e1e2}), the breakup events group themselves in a prominent 
peak, underlaid with only a low background of accidental coincidences. 
As it has been already pointed out in~\cite{Kis03a}, the choice of
integration limits $D_{\textrm{a}}$ and $D_{\textrm{b}}$, as well as of 
the assumed background function, is not critical since the contribution 
of accidental coincidences in all analyzed angular configurations was 
very low (between 2\% and 5\%). However, to treat all configurations in 
a consistent way, and since all the $D$-projected distributions have 
approximately Gaussian shape, the limits $D_{\textrm{a}}$ and 
$D_{\textrm{b}}$ were chosen at the values of $-3\sigma$ and $+3\sigma$ 
from the maximum of the fitted peak. 
A linear dependence of background between those points was assumed.
Gaussian shape and linear background fitted to a sample distribution 
are shown in the inset of Fig.~\ref{fig_e1e2}.

\subsubsection{\label{secIIB2}Detection efficiency}

\begin{figure}
\includegraphics[width=80mm]{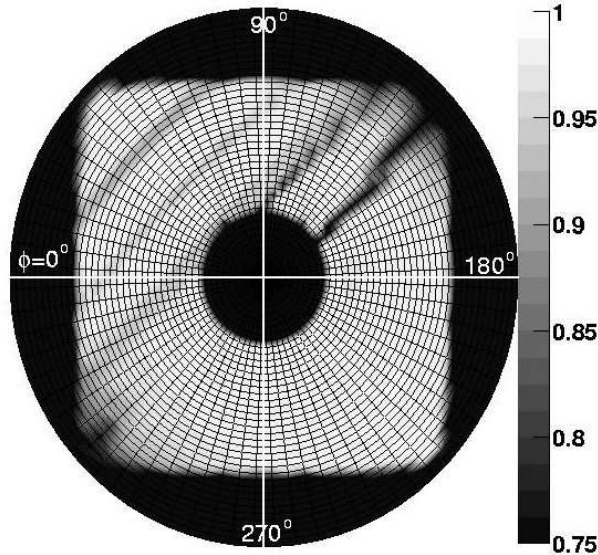}
\caption{\label{fig_effmwpc}
Global efficiency map for the MWPC, presented as a polar plot in
the angular coordinates.  Only the square-like area of the map is
meaningful. 
The range of polar angle $\theta$ is from 0$^{\circ}$ to 40$^{\circ}$ 
with the binning of 1$^\circ$ and the azimuthal angle $\phi$ covers the
full range with the binning of 5$^\circ$.  The discontinuities of lower
efficiency regions are the result of this finite binning.}
\end{figure}

The efficiency in determining the particle-emission angles, called for 
simplicity the MWPC efficiency, is a product of hardware efficiencies 
of the MWPC wire planes and the efficiency of our procedure of 
reconstructing the angles.  Since we accepted only events with vertical 
and horizontal wires properly correlated with the corresponding $E$ and 
$\Delta E$ detectors, the ranges of wire numbers associated with the 
individual hodoscope elements had to be set wide enough.  These 
correlation tables were revised once again (with respect to the 
procedure of Ref.~\cite{Kis03a}), inspecting the whole data sample 
and the ranges of wires associated with each hodoscope element have 
been slightly broadened. In this way, since there was 
practically no uncorrelated noise on the wires, no additional background
was introduced while the efficiency was increased.  With these new 
conditions the efficiency  has been recalculated in the manner similar 
to the one described in~\cite{Kis03a} and, in addition, losses due to 
the requirement of correlation between all three planes have been 
determined.  In spite of imposing this last restriction with 
reasonable ``safety limits'' of $\pm 3$  wires in $u$ plane, some protons
scattered on their way to MWPC escaped those limits and were rejected, 
affecting the total efficiency. The final map of the MWPC 
efficiency is presented in Fig.~\ref{fig_effmwpc}. 

\begin{figure}
\includegraphics[width=86mm]{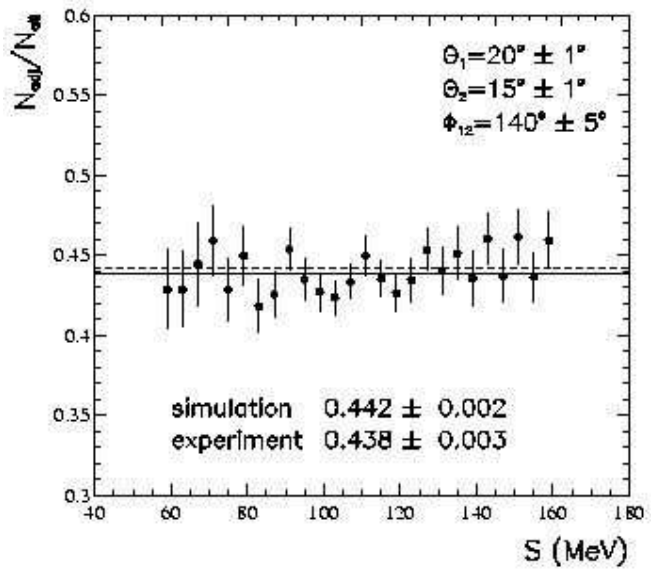}
\caption{\label{fig_effsec1}
Ratio of event rate of proton-proton coincidences registered in two 
adjacent sectors to the total rate of events (adjacent + diagonal) 
for configuration $\theta_1 = 20^{\circ} \pm 1^{\circ}$, 
$\theta_2 = 15^{\circ} \pm 1^{\circ}$, 
$\phi_{12} = 140^{\circ} \pm 5^{\circ}$.
Points represent experimental values obtained by integrating the events
within the given bin of the arclength $S$.  Solid line shows the
level fitted to the data, the dashed one the ratio obtained from
the simulation.  Numerical results with statistical errors are given
for both, experimental and simulated data.}
\end{figure}

Thorough studies of the detection and trigger efficiencies were 
supplemented with additional tests, performed for configurations with 
the relative azimuthal angle of the two protons exceeding 90$^{\circ}$.  
Such configurations can be realized by two mutually exclusive classes 
of events: when the two protons were registered in either the adjacent 
sectors or in two diagonal sectors of the hodoscope.  These two cases 
corresponded to different trigger signals and different downscaling 
factors, therefore any relative inefficiency of the trigger logic
and/or of the downscaling should be reflected by influencing the 
relative amount of events of the two groups. 
It should be stressed that we are sensitive only to the {\em relative}\/
efficiency of the two trigger classes: since the events originating 
in the elastic scattering are always registered in diagonal sectors, 
the trigger/detection efficiency for the diagonal sectors cancels out 
in normalization, cf.~Eq.~(\ref{br_cs2}). 
Each of the two event groups was analyzed separately and the ratio of 
their rates as a function of the arclength $S$ was constructed.  
An analogous ratio was calculated  for simulated events. For all the 
configurations the experimental ratio is constant along $S$ and agrees 
with the result of simulations within statistical accuracy of 0.8\% or less,  
depending on the configuration (see example in Fig.~\ref{fig_effsec1}).  
The above result confirms not only the correct functioning of the trigger
but also proper handling of the detection efficiencies in the analysis.

\begin{figure}

\includegraphics[width=86mm]{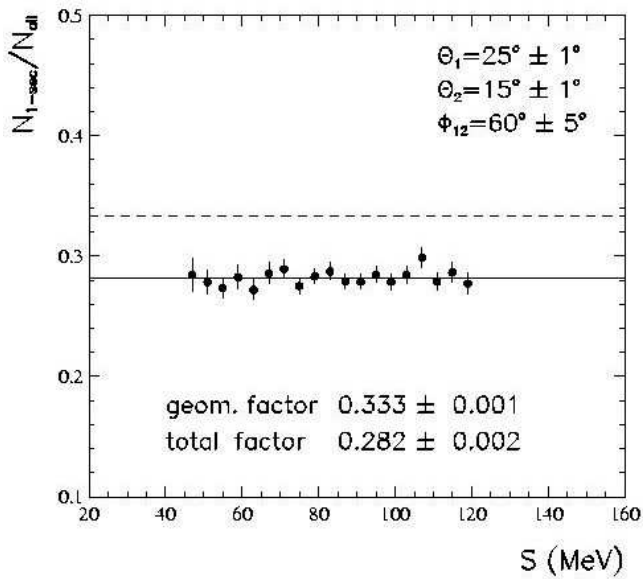}
\caption{\label{fig_eff3}
Ratio of the rate of single-sector events to the total 
rate of events (one-sector plus two-sector) 
for configuration $\theta_1 = 25^{\circ} \pm 1^{\circ}$, 
$\theta_2 = 15^{\circ} \pm 1^{\circ}$, 
$\phi_{12} = 60^{\circ} \pm 5^{\circ}$.
Points show the experimental values obtained within 4 MeV $S$-bins. 
The difference between the purely geometrical factor (dashed line)
and the total correction factor (solid line) including the acceptance 
losses for single-sector events, is clearly visible.
Numerical values of the factors are given with their statistical 
uncertainties.}
\end{figure}

In configurations with the relative azimuthal angle of two protons 
$\phi_{12}$ not exceeding 90$^{\circ}$ additional losses of acceptance 
have to be taken into account, due to cases when both protons were 
registered in the same $\Delta E$ or $E$ detector (impossible proper 
particle identification and/or energy determination), or in two adjacent 
$E$ detectors.  In the latter case, if at least one of the two protons was 
registered close to the edge of the two $E$ detectors, the event cannot 
be distinguished from a so-called cross-over and had to be rejected 
(cf.~Sec.~\ref{secIIB3}).  Obviously, only single-sector events are 
affected by these effects.  Therefore, the total correction factor is 
obtained as a product of the ratio of events with both protons emitted 
into a single sector to all events collected in the specific configuration, 
and the actual correction, describing the losses within the single sector.  
Experimentally, the total correction can be determined by the ratio of the 
breakup events {\em registered} in single sectors to all events of the 
considered configuration.  By means of simulation one can investigate both
contributions separately.  First, in order to find for each configuration
with $\phi_{12} < 100^{\circ}$ purely geometrical factors of probability
for single-sector events, an ideal case was assumed with no losses due to 
the detector granularity.  Then, the events were artificially
digitized and analyzed in the same way as the experimental ones.
It has been found that the acceptance losses reduce the geometrical
factors by up to 13\%, depending on the configuration, but they 
are constant as a function of $S$ for the selected geometry.  
In Fig.~\ref{fig_eff3} an example comparing the pure geometrical
and the total correction factors is shown for one configuration.  
In general the losses increase with decreasing relative azimuthal 
angle and with the increasing difference of the polar angles.  It should 
be noted that the errors introduced by applying the correction 
factors are much smaller than the factors themselves.
    
\subsubsection{\label{secIIB3}Cross-over correction}

The procedure of energy calibration faces a problem in determining the 
energy of particles which penetrate from one stopping detector
to the adjacent one, in the so called cross-over events. Simple summing up 
of the two deposited  energies is not completely adequate due to energy
losses in the foil covering all the detector walls.  Additionally, in a 
particular situation when the energy deposited in one of the $E$ slabs is 
below the detection threshold, the energy information is significantly 
distorted and, moreover, there is even no obvious signature of cross-over. 
Such events are shifted away from the kinematic curve and contribute to
the background attributed to accidental coincidences, which is then 
subtracted in the way described in Section~\ref{secIIB1}. 

\begin{figure}
\includegraphics[width=86mm]{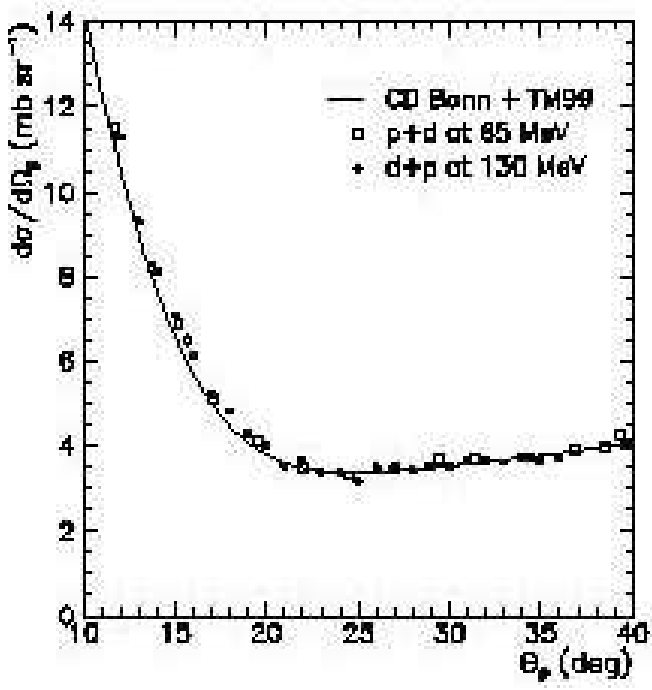}
\caption{\label{fig_csel}
Experimental angular distribution of the elastic scattering events.
The absolute normalization of our data (full circles) is adjusted
to best fit the reference data.  Statistical errors are smaller 
than the point size.  Empty squares represent the reference 
cross-section values~\cite{Shi82a}. Solid line shows the results of 
the theoretical calculations obtained with the CD Bonn potential and 
the TM99 3NF model.}
\end{figure}

Therefore, a new approach was used, in which all cross-over candidates
were rejected from the analysis (in a way explained below) and their 
amount was determined with the use of Monte-Carlo simulation based 
on the {\tt GEANT4} package.  Narrow regions
corresponding to the detector borders were defined with the help of
high-resolution MWPC position coordinates and the particles which entered
those regions and induced signals in two adjacent $E$ detectors 
were discarded.  Treating the simulated data in the same way it was 
possible to find the ratio of rejected to registered event
numbers for every configuration, which was used to correct the experimental
rates.  The simulations were performed for elastic scattering and 
for all studied configurations of the breakup reaction. For elastic 
scattering, due to rather high proton energy and constrained kinematics 
the effects are quite large -- on average about 7\% of all the events is
biased with the cross-over possibility.  The corresponding correction
factors vary strongly, from 4\% to 11\%, depending on the proton polar 
angle $\theta_{p}$.  Their impact is demonstrated by the fact that
application of this correction leads to the experimental cross-section 
distribution for elastic scattering (Fig.~\ref{fig_csel}) with a 
smoother dependence on $\theta_{p}$, following more closely the 
reference $pd$ data.  It is also reflected by a decrease of the $\chi^2$
value calculated between the two distributions by a factor of about 2 
with respect to the result obtained for the uncorrected data. 
Contributions of the cross-over events, calculated individually 
for each analyzed configuration of the breakup reaction, vary between 
2\% and 5\%.  The individual cross-over correction factors were applied 
when evaluating the differential breakup cross sections, resulting in a
decrease of their systematic uncertainty.

\subsubsection{\label{secIIB4}Cross section normalization}

The breakup cross sections are normalized to the elastic scattering one,
using the measured in parallel rates of the elastic scattering events and 
the available $pd$ elastic scattering cross-section data. 
The differential breakup cross section for a chosen angular configuration 
is thus expressed in terms of the elastic scattering cross section and both 
measured coincidence rates: 
\begin{eqnarray}
 \frac{d^{5}\sigma}{d\Omega_{1}d\Omega_{2}dS}(S,\Omega_{1},\Omega_{2}) = 
 \frac{d\sigma_{el}}{d\Omega_{1}^{el}}(\Omega_{1}^{el}) \cdot 
\frac{N_{br}(S,\Omega_{1},\Omega_{2})}{N_{el}(\Omega_{1}^{el})} \times 
 \nonumber \\
 \times \: \frac{\Delta \Omega_1^{el}}
{\Delta \Omega_{1} \Delta\Omega_{2} \Delta S}
\cdot \frac{\epsilon^{el}(\Omega_{1}^{el})\epsilon^{el}(\Omega_{2}^{el})}{
\epsilon(\Omega_{1})\epsilon(\Omega_{2})}\;,\;\, 
\label{br_cs2}
\end{eqnarray}
where $N_{br}$ is the number of breakup coincidences registered at the 
angles $\Omega_{1}$, $\Omega_{2}$ and projected onto a $\Delta S$-wide 
arclength bin.  Subscripts 1 and 2 refer to the first and the second 
proton registered in coincidence or to the proton and the deuteron in 
the case of elastic scattering.
$\Omega_{i} \equiv (\theta_{i},\phi_{i})$, with $i=1,2$, are the polar and 
azimuthal angles, respectively, and $\Delta\Omega_i$ is the solid angle 
($\Delta\Omega_{i}=\Delta\theta_{i}\Delta\phi_{i}\sin{\theta_{i}}$). 
Products $\epsilon(\Omega_{1})\epsilon(\Omega_{2})$ 
(or $\epsilon^{el}(\Omega_{1}^{el})\epsilon^{el}(\Omega_{2}^{el})$) 
contain all relevant efficiencies and correction factors 
(cf.~Sec.~\ref{secIIB3}). 
$N_{el}$ is the final number of elastic scattering $pd$ coincidences 
registered at the proton angle $\Omega_1^{el}$.  The elastic scattering 
cross section $\frac{d\sigma_{el}}{d\Omega_{1}^{el}}(\Omega_{1}^{el})$ 
is taken from~\cite{Shi82a}. The bin width $\Delta S$ was chosen to be 
4~MeV. 

In such an approach we profit from cancellation of all factors related
to the luminosity, i.e.~the integrated beam current, the density and 
the thickness of the target. Moreover, since events from both reactions 
are processed by common electronic and read-out systems, the relevant
dead-time corrections cancel out in the ratio of the registered events.
In that way factors which would be difficult to determine individually
and would induce systematic uncertainties are greatly eliminated.  

\subsubsection{\label{secIIB5}Experimental uncertainties}

A full discussion of the experimental uncertainties has been presented 
in~\cite{Kis03a}.  The statistical accuracy comprises the error of 
the measured number of the breakup coincidences, as well as statistical 
uncertainties of all quantities used in the cross-section normalization, 
i.e.~the number of the elastic scattering events and all efficiencies 
included in Eq.~(\ref{br_cs2}).  Taking into account the range of the 
cross-section values for our data points, the magnitude of the statistical 
errors varies between 0.5\% and 4.0\%.
 
Non-negligible systematic effects can originate from the cross section
normalization, from uncertainty of the energy calibration parameters, 
from incomplete cancellation of polarization effects 
in the cross section (1.0\%) and from the procedure of reconstruction 
of the proton emission angles.  By introducing the cross-over corrections 
we were able to suppress variations in the experimentally obtained 
elastic scattering distribution and therefore reduce the total uncertainty 
of the normalization procedure down to about 2.0\%.  The uncertainty
of the energy calibration can result in changing the length of the
experimental distribution along the $S$-curve by at most 0.7\%.  
The relative cross-section errors resulting from such change vary 
between 0.7\% and 2.5\%, for central and peripheral regions of the 
measured $S$-ranges.  This is the only systematic uncertainty which 
changes along the arclength $S$ in every configuration; all other 
contributions are rather configuration-specific. 

The uncertainty of the reconstructed value of the angle is due mainly 
to finite target thickness, finite size of the beam 
spot on the target, straggling effects and angular resolution related 
to the discreteness of the position information delivered by the MWPC.  
The well-reproduced correlation of the proton and deuteron emission 
angles for elastic $dp$ scattering confirms that there is no (at the 
level below 0.3$^{\circ}$) systematic shift of the reconstructed polar 
angles. 
The other effects, resulting in smearing out of the angular resolution, 
have been studied by the dedicated simulation, based again on the 
{\tt GEANT4} package.  In order to reproduce conditions of 
the real measurement,  realistic distribution of the reaction vertices and
theoretical angular dependence of the breakup cross section were assumed.  
Straggling effects in materials were introduced by means of {\tt GEANT4} 
transport routines and, finally, the positions of proton trajectories
intersecting the wire planes were translated to hits on the wires. 
The same reconstruction algorithm as for the real data was applied to 
the simulated events. In this way we were able to compare the amount of 
events in each configuration defined by proton emission angles, with the 
number of events in the same configurations, but defined with the use of 
the reconstructed angles.  It was found that for about 30\% of events 
selected on the basis of the reconstructed angles the particles were 
really emitted at angles lying outside the chosen angular range.  On the 
other hand, practically the same amount of events emitted into the chosen 
range is reconstructed with the values of angles not belonging to the
considered configuration, and therefore rejected.  In this way the number 
of events in ``true'' and ``reconstructed'' configurations is very well 
balanced: differences are between 0.2\% and 1.0\% and do not contribute
significantly to the cross section errors. 

\begin{table}
\caption{\label{tab_err}
Summary of the relevant experimental cross-section uncertainties.
Two sample cross-section data points (with values close to the minimal
and maximal ones measured) are selected for presenting individual
contributions to the systematic uncertainties:\\
{\bf (1)} $\theta_1 = \theta_2 = 15^{\circ}$, $\phi_{12} = 60^{\circ}$,
$S$ = 106 MeV,\\ \hspace*{5.2 mm}
d$^5\sigma$/d$\Omega_1$d$\Omega_2$d$S$~=~0.078~mb$\cdot$sr$^{-2}\cdot$MeV$^{-1}$,\\
{\bf (2)} $\theta_1 = \theta_2 = 25^{\circ}$, $\phi_{12} = 160^{\circ}$,
$S$ = 134 MeV,\\ \hspace*{5.2 mm}  
d$^5\sigma$/d$\Omega_1$d$\Omega_2$d$S$~=~1.57~mb$\cdot$sr$^{-2}\cdot$MeV$^{-1}$.\\
The last column shows all the overall ranges of the relative cross section
uncertainties.  The ``total systematic'' error is obtained by adding
the squares of all the contributions.}
\begin{ruledtabular}
\begin{tabular}{lccc}
~ & $\delta \sigma_1$ & $\delta \sigma_2$ & $\delta \sigma$ range \\
Source of uncertainty & (\%) & (\%) & (\%) \\
\hline
{\bf Statistical}            & 2.7 & 0.6 & {\bf 0.5 -- 4.0} \\
~\\
Energy calibration           & 1.9 & 0.7 & 0.7 -- 2.5 \\
Beam polarization            & 1.0 & 1.0 & 1.0 \\
Reconstruction of angles     & 0.6 & 0.5 & 0.2 -- 1.0 \\
Choice of integration region & 0.3 & 0.1 & 0.1 -- 1.0 \\
Normalization:               & 1.6 & 2.0 & 1.6 -- 2.0 \\
{\bf Total systematic}       & 2.8 & 2.4 & {\bf 2.0 -- 3.6} \\
\end{tabular}
\end{ruledtabular}
\end{table}

The complete simulation of the breakup process lead to the conclusion that
the influence of angular resolution on the cross section is in fact smaller
than what was found from the geometrical estimations~\cite{Kis03a}.  
Including all improvements of the current analysis, the total systematic 
uncertainty is lower by about 1\% as compared to the previously quoted 
values.  The experimental uncertainties relevant for the cross sections 
presented here are summarized in Table~\ref{tab_err}.  The overall ranges 
of uncertainties (last column) are not to be associated with particular
magnitudes of the cross sections.  An obvious exception is apparently
the statistical accuracy; however, since data were collected with different
downscaling factors and there are certain acceptance losses, this 
scaling is also not straightforward.  Therefore, we have selected two 
cross-section points with values close to the minimal and maximal of 
all presented in Sec.~\ref{secIV} and we display for them the individually 
calculated contributions to their uncertainties.  
One can observe that uncertainties of the larger measured cross sections 
are usually dominated by systematic effects, while for the smaller values
of cross sections the contributions from systematic and statistical
errors are comparable.

\section{\label{secIII}Theoretical Formalism}

\subsection{\label{secIIIA}Realistic potentials}

The calculation of the cross-section values using realistic
potentials is performed exactly as outlined in our previous 
study~\cite{Kis03a}, following our standard method for the 3$N$ 
continuum.
The general overviews of our formulation of the 3$N$ scattering 
problem and of including 3NF into the scheme are given 
in~\cite{Glo96a} and~\cite{Hub97a}, respectively.  In the following 
a very brief review is presented.

We use the modern, realistic $NN$ potentials AV18~\cite{AV18}, charge 
dependent (CD)~Bonn~\cite{CDBonn1,CDBonn2}, and Nijm I and 
II~\cite{NijmIII}.  Investigating the full 3$N$ system dynamics, we 
combine them with the 2$\pi$-exchange TM 3NF, taking its recent 
form~\cite{TM993NF}, consistent with chiral symmetry, which will be
denoted by TM99.  The TM99 3NF model contains one parameter, 
$\Lambda_{TM}$, used as cut-off to regularize its high-momentum 
behavior.  The value of $\Lambda_{TM}$ is adjusted for each particular 
combination of the $NN$ force and the TM99 3NF to match the value 
of the $^3$H binding energy~\cite{Nog97a}.  For the four 2$N$ potentials 
used in the calculations the corresponding values of $\Lambda_{TM}$ 
(in units of the pion mass $m_{\pi}$) are 4.764, 4.469, 4.690 and 4.704, 
respectively.

When the 3$N$ system dynamics is studied with the AV18 $NN$ potential, we 
combine it also with the Urbana~IX 3NF~\cite{UIX3NF} (UIX).  To apply
it within our framework it was necessary to transform its
configuration-space form to momentum space~\cite{Wit01a}. 

Having the $NN$ and 3$N$ forces, the scattering problem in the 3$N$
system is stated in form of a Faddeev-like integral equation for an 
amplitude $T$:
\begin{eqnarray}
T & = & t \, P \, \phi \ 
+ \ ( 1 + t G_0 ) \, V_{3NF}^{[1]} \, ( 1 + P ) \, \phi \ 
+ \ t \, P \, G_0 \, T \nonumber \\
 & & \mbox{} + \ ( 1 + t G_0 ) \, V_{3NF}^{[1]} \, ( 1 + P ) \, G_0 \, T\, ,
\label{eqFt}
\end{eqnarray}
where the initial channel state $\phi$ is composed of a deuteron and a
momentum eigenstate of the projectile nucleon. The $NN$ transition operator 
is denoted by $t$, the free 3$N$ propagator by $G_0$ and $P$ is the sum of 
a cyclical and an anti-cyclical permutation of the three particles. 
The 3$N$ potential $V_{3NF}$ can always be decomposed into a sum of 
three parts:
\begin{equation}
V_{3NF} = V_{3NF}^{[1]} + V_{3NF}^{[2]} + V_{3NF}^{[3]}\, ,
\label{eqF3NF}
\end{equation}
where the part $V_{3NF}^{[i]}$ singles out nucleon $i$, on which the
pion is rescattered.  The parts are symmetric under the exchange of
the two nucleons $j$ and $k$, with $j \ne  i \ne k$.
One can see that in Eq.~(\ref{eqFt}) only one part, $V_{3NF}^{[1]}$ 
appears explicitly; the others enter via the permutations contained 
in $P$. The physical breakup amplitude $U_0$ is obtained from $T$ by 
\begin{equation}
U_0 = ( 1+ P) T\, .
\label{eqFu0}
\end{equation}

Iterating the Faddeev-like equation~(\ref{eqFt}) and inserting the 
resulting $T$ into Eq.~(\ref{eqFu0}) yields the multiple scattering 
series, in which each term contains some number of interactions 
among nucleons via 2$N$- and 3$N$-forces with free propagation 
in between. The reaction mechanism is thus transparently mirrored.

We solve Eq.~(\ref{eqFt}) using a momentum space partial-wave 
basis~\cite{Glo96a}. To guarantee converged solutions for our
case of 130 MeV incoming deuterons we take into account all
partial waves with $j_{max} < 6$ in the 2$N$ subsystem. This
gives rise to the maximal number of 142 partial wave states 
in the 3$N$ system for each total 3$N$ angular momentum $J$.
The convergence has been checked by inspecting the results 
obtained for $j_{max} = 6$ calculations without a 3NF (total
number of channels increased to 194). Finally, the breakup 
amplitudes $U_0$ have been calculated for all total angular 
momenta of the 3$N$ system up to $J = 25/2$ for any $NN$ 
interaction, while the inclusion of 3NF's has been carried 
out for all states up to $J = 13/2$.  From this amplitude
the cross section is obtained in a standard manner~\cite{Glo96a}.

\subsection{\label{secIIIB}Chiral Perturbation Theory}

The chiral 2$N$ potential at the next-to-next-to-leading order (NNLO) used 
in the present study is derived from the most general effective chiral 
Lagrangian, based on the method of unitary transformation~\cite{Epe98a} 
and using the spectral function regularization (SFR)~\cite{Epe05a}. 
More details about the employed regularization schemes and the 
corresponding cut--offs can be found in~\cite{Epe05a}.
Completing the 3$N$ system dynamics at NNLO with the naturally and 
consistently arising 3NF contributions is presented in~\cite{Epe02a}.  
We recall below a few key features.

The 2$N$ force is obtained by summing up contributions from graphs
of increasing complexity, accounting for, roughly speaking, two kinds
of processes: long range pion(s) exchanges, where chiral symmetry 
plays a crucial role and short range phenomena, which are effectively
treated by means of $NN$ contact interactions. The corresponding 
low energy constants (LEC's) are determined from the $NN$ data. 
The potential is expressed in terms of the expansion in powers $\nu$ of
$Q/\Lambda_{\chi}$, where $Q$ is the soft scale, corresponding to the 
nucleon external momenta and the pion mass and $\Lambda_{\chi}$ is the 
hard scale (around 1 GeV) associated with the chiral symmetry breaking 
scale or the ultraviolet cut--off(s).  For each diagram contributing 
to the potential, the power $\nu$ can be calculated according to the 
power counting scheme~\cite{Wei90a}, see also~\cite{Epe98a}. 

Up to NNLO, the chiral $NN$ potential can be written  as
\begin{equation}
V^{NN} = V^{(0)} + V^{(2)} + V^{(3)}\, .
\label{eqChV2N}
\end{equation}
At the leading order (LO, $\nu$ = 0) the $NN$ potential $V^{(0)}$ is
given by the one-pion exchange part (1PE) and two contact interactions:
\begin{equation}
V^{(0)} = V_{1\pi}^{(0)} +  V_{cont}^{(0)}\, .
\label{eqChV0}
\end{equation}
The leading 1PE term is expressed in terms of standard constants:
the pion decay constant $F_{\pi}$, the pion mass $m_{\pi}$ and the 
axial-vector nucleon coupling $g_{A}$. In the contact term two LEC's 
are introduced, $C_{S}$ and $C_{T}$.  The next-to-leading order (NLO,
$\nu$ = 2) corrections are due to two-pion exchanges (2PE), seven
new contact interactions and a correction to 1PE:
\begin{equation}
V^{(2)} = V_{2\pi}^{(2)} + V_{1\pi}^{(2)} +  V_{cont}^{(2)}\, .
\label{eqChV2}
\end{equation}
The leading 2PE term introduces no new parameters (except the SFR cut-off,
see below), the contact terms are characterized by seven constants
$C_1, ..., C_7$ and in the 1PE correction term the constant $d_{18}$ can 
be incorporated by renormalization of $g_{A}$ -- see~\cite{Epe05a,Epe03a} 
for more details.  Finally, the NNLO ($\nu$ = 3) corrections are given by 
the subleading 2PE potential and corrections to the 1PE force:
\begin{equation}
V^{(3)} = V_{2\pi}^{(3)} + V_{1\pi}^{(3)}\, ;
\label{eqChV3}
\end{equation}
there are no new contact terms.  The 2PE term contains three new
LEC's, $c_1$, $c_3$ and $c_4$.  
The LEC's $C_S$, $C_T$ and $C_1 , \ldots , C_7$, appearing at LO and NLO,
are obtained by fitting the $V^{NN}$ predictions to the $NN$ data (more
precisely, to the lowest phase shifts).  The three LEC's $c_1$, $c_3$ 
and $c_4$ entering the 2PE contribution at NNLO can be determined from 
the $\pi N$ scattering data.  It has been shown~\cite{Epe04a} that the 
adopted values lead to a proper reproduction of the deuteron properties 
and of the phase-shift analysis results.

A non-vanishing chiral 3NF arises only at NNLO (in the energy-independent 
formulation) and can be written as
\begin{equation}
V^{3NF} = V^{3NF}_{2\pi} + V^{3NF}_{1\pi} + V^{3NF}_{cont}\, .
\label{eqChV3N}
\end{equation}
The three terms account for three different topologies~\cite{Epe02a}.  
The $V^{3NF}_{2\pi}$ describes a simultaneous exchange of two pions
and incorporates the same LEC's $c_{1,3,4}$ as in the subleading
2PE $NN$ potential $V_{2\pi}^{(3)}$.  The $V^{3NF}_{1\pi}$ contribution
arises due to a single pion being exchanged between a nucleon and a 2$N$ 
contact interaction.  The contact term contains one parameter, which is 
usually called $c_{D}$.  Finally, the $V^{3NF}_{cont}$ describes a contact 
interaction of three nucleons and introduces another LEC, labeled $c_{E}$.
These two last LEC's are fixed by the requirement to reproduce the
$^3$H binding energy and the $nd$ doublet scattering length -- 
see~\cite{Epe02a} for a detailed discussion.

In representing the chiral potential we use the spectral function 
decomposition~\cite{Epe04a} and we reject the large-mass (momentum)
fraction of the 2PE via a step Heaviside function with
the cut-off parameter $\tilde{\Lambda}$.  In the 2PE contributions
at NLO and NNLO the loop functions are thus regularized and the 
corresponding short-distance phenomena are shifted into the
contact terms (the LEC's are appropriately adjusted).  This 
procedure, called SFR, 
possesses numerous advantages over formerly implemented dimensional 
regularization~\cite{Epe04a,Epe05a}. Its implementation allowed us 
also to use LEC's $c_{1,3,4}$ consistent with the $\pi N$ data, in
contrast to the former study~\cite{Epe02a}, where the 3$N$ dynamics
was described in a so-called NNLO* approach, with artificially small 
values of these constants.

Using the resulting potential, the $t$-matrix is obtained via numerical, 
non-perturbative solution of the partial-wave projected Lippmann-Schwinger 
(LS) equation.  
Since the effective 2$N$ forces are meaningless for large momenta, 
we still have to reject contributions of the high-momentum states.  
In this way we also avoid an ultraviolet divergence of the LS equation. 
The standard procedure to accomplish those requirements is to
regularize the potential by multiplying it with a regulator function,
containing an additional cut-off parameter $\Lambda$. As in other
studies~\cite{Epe04a,Epe05a}, we use a Gaussian regulator function.
The $T$ operator of the 3$N$ scattering problem is obtained in
an identical way as for realistic potentials (cf.\ Sec.~\ref{secIIA}), 
by solving the Faddeev-like equation (Eq.~\ref{eqFt}) with the chiral 
3NF from Eq.~(\ref{eqChV3N}). To keep the treatment of 2$N$ and 3$N$ 
interactions consistent, we use an appropriate regulator function with 
the same cut-off parameter $\Lambda$ as for the $NN$ potential in the 
LS equation also for regularization of the chiral 3NF. In further 
calculations the observables are generated on the basis of the obtained 
breakup amplitude $U_0$.

Our method provides the possibility to estimate uncertainties of the
calculated predictions. We perform calculations with a few 
combinations of the two cut-off parameters, [$\Lambda$,$\tilde{\Lambda}$].
The range of predictions obtained for reasonable choice of the
variation intervals of both cut-offs gives an estimate of the 
theoretical uncertainty. For details on how one selects the proper
ranges of regularization cut-off values we refer to~\cite{Epe04a,Epe05a} 
and references therein. In the present study we use the following
pairs of the cut-off parameters (values in MeV):
\begin{eqnarray}
[\Lambda,\tilde{\Lambda}] & = & [450,500], [600,500], [550,600], \nonumber \\
 & & [450,700], [600,700]\, .
\label{eqChLmbd}
\end{eqnarray}

\subsection{\label{secIIIC}Coupled-channel potential}

A new realistic two-baryon potential coupling $NN$ and $N\Delta$ states
has been presented in detail in~\cite{Del03b}, with several examples
of its application to calculate observables for the 3$N$ system.
The main features of this approach are briefly recalled here.

The dynamics of the 3$N$ system is described with the explicit treatment
of the $\Delta$-isobar excitation, considered in the relevant energy
range as a stable particle.  The three nucleon channels are coupled
to those in which one nucleon is excited and forms the $\Delta$-isobar.
Creation of a virtual excited state yields an effective 3NF, in parallel
to other aspects of the dynamics induced by the $\Delta$-isobar.

The method for extending a model of $NN$ interaction to include 
$\Delta$-isobar degrees of freedom has been worked out in~\cite{Haj83a}
and recently thoroughly upgraded~\cite{Del03b}, taking the purely 
nucleonic CD Bonn potential~\cite{CDBonn2} as a reference.
Such a coupled-channel potential is based on the exchange of $\pi$,
$\rho$, $\omega$, and $\sigma$ mesons, and in addition to the purely 
nucleonic part  
includes also contributions from the transition between the $NN$ and
$N\Delta$ states, from the exchange $N\Delta - \Delta N$ potential
and from the direct interaction of the $N\Delta$ states.  The force 
employed here, referred to as CD Bonn~+~$\Delta$, is as realistic as 
any of the $NN$ force models quoted in Sections~\ref{secI} 
and~\ref{secIIIA}, reproducing the data of the 2$N$ system with a 
$\chi^2$ per degree of freedom of 1.02~\cite{Del03b}.  
It is purely nucleonic in the isospin singlet states; the coupled-channel 
two-baryon extension acts in isospin triplet states only, where a few 
constants of the reference $NN$ force are retuned.  Prominent contributions 
of the effective 3NF mediated by the $\Delta$-isobar are of the 
Fujita-Miyazawa type~\cite{FM3NF} and of the Illinois ring 
type~\cite{Pie01a}.
The contributions are based on all meson exchanges, i.e.\ $\pi$, $\rho$,
$\sigma$ and $\omega$ exchanges, contained in the coupled-channel 
potential; the $\Delta$ propagation is retarded.
The arising effective three-nucleon force is much richer with respect
to the $\Delta$ excitation and also has
shorter ranged components than standard irreducible two-pion exchange
3NF's. Furthermore, all its components are dynamically
consistent with each other and with the effective 2$N$ force.
However, in addition to the $\Delta$-mediated 3NF an irreducible
3NF covering other physics mechanisms is \emph{not} used.

The solution of the three-baryon scattering problem is based on the 
AGS equation formulation, using a Chebyshev expansion of the 
two-baryon transition matrix as the interpolation technique~\cite{Del03a}.
The multichannel transition matrix $U$ between two-body channels 
is obtained from:
\begin{equation}
U = P \, G_0^{-1} +  P \, T_{\alpha} \, G_0 \, U  \, ,
\label{eqCCu}
\end{equation}
where $T_{\alpha}$ is the two-baryon transition matrix in three-baryon space
(the subscript $\alpha$ denoting the pair $\beta$-$\gamma$ of interacting 
baryons, $\alpha \ne \beta \ne \gamma$), $G_0$ is the free resolvent 
$(E - i0 - H_0)^{-1}$ with the total available energy $E$ and free 
Hamiltonian $H_0$, and $P$ is the permutation operator introduced in 
Eq.~(\ref{eqFt}). The transition matrix $T_{\alpha}$ results from
the full form of the two-baryon potential $V_{\alpha}$, acting between
baryons $\beta$ and $\gamma$:
\begin{equation}
T_{\alpha} = V_{\alpha} +  V_{\alpha} \, G_0 \, T_{\alpha} \, .
\label{eqCCv}
\end{equation}
The breakup transition matrix $U_0$ is obtained from $U$ according to
\begin{equation}
U_0 = ( 1 + P) \, G_0^{-1} +  ( 1 + P) \, T_{\alpha} \, G_0 \, U  \, .
\label{eqCCu0}
\end{equation}
The first term on the right side of Eq.~(\ref{eqCCu0}) does not
contribute to the on-shell matrix elements of $U_0$, needed to calculate
breakup observables.  The approach described here is very
similar to the one outlined in Sec.~\ref{secIIIA}.  If we define
the amplitude $T$ as:
\begin{equation}
T = T_{\alpha} \, G_0 \, U  \, ,
\label{eqCCt}
\end{equation}
then the integral AGS equation~(\ref{eqCCu}) after simple algebra becomes
identical with the Faddeev-like equation~(\ref{eqFt}), in which the
3NF potential is set to zero and the $NN$ $t$-operator is identified
with $T_{\alpha}$.  Following the remark below Eq.~(\ref{eqCCu0}),
that equation becomes immediately identical with Eq.~(\ref{eqFu0}).

Matrix elements of the amplitudes $U$ and $T$, necessary to calculate
breakup observables, are found in the partial-wave basis.  The charge
dependence of the two-baryon potential is treated as described 
in~\cite{Del03b}, yielding the total isospin $\frac{3}{2}$ channels
in the $^1S_0$ state. 
In the purely nucleonic channels all the states with $j_{max} < 6$ in the
two-baryon system have been taken into account, while for the $N\Delta$ 
channels the applied total angular momentum limitation was $j_{max} < 5$.
In the full three-baryon space the states with angular momentum up to
$J_{max} = 31/2$ were taken into account.  For the energy considered here, 
the results are fully convergent with respect to both, $j_{max}$ and $J_{max}$ 
limitations, what was tested by checking several predictions obtained 
with the limits $j_{max} = 6$ and $J_{max} = 35/2$.

The discussion of the coupled-channel potential approach should be
closed with a few remarks.  The mechanism of explicit $\Delta$ excitation
in the three-baryon interaction has two distinct effects: it yields
an effective repulsive potential (two-baryon dispersion) and it 
induces an effective 3NF.  These two contributions usually 
compete~\cite{Del03b,Del03c}, resulting in relatively modest 
differences when comparing the results of CD Bonn and 
CD Bonn~+~$\Delta$ predictions.  The competition might be less 
pronounced at higher energies.  It should be also noted that in this
method the binding energies of the 3$N$ systems are reproduced 
a bit less perfectly than in the other approaches.  On the other
hand, only in this framework has a significant development towards 
including the Coulomb interaction into the Faddeev formalism for the
3$N$ continuum been achieved recently~\cite{Del05a}.

\section{\label{secIV}Results}

The main purpose of this paper is a {\em systematic}\/ study of the
quality by which the breakup cross sections can be reproduced by
theoretical predictions.  The investigation spans a significant
fraction of the breakup reaction phase space, the attainable 
geometries defined by our experimental conditions.  In our 
methodical approach of scanning the phase space we present the 
cross-section data for a regular grid of polar and azimuthal angles
with a constant step in the arclength variable $S$.   
Polar angles of the two protons $\theta_1$ and $\theta_2$ are changed 
between 15$^{\circ}$ and 30$^{\circ}$ with the step size of 5$^{\circ}$ 
and their relative azimuthal angle $\phi_{12}$ is analyzed in the range 
from 40$^{\circ}$ to 180$^{\circ}$, with the step size of 20$^{\circ}$.
We are able to extract data covering a denser grid. However, 
since the changes of the breakup cross section are rather smooth,
already this coverage allows to draw all the important conclusions.  
For each combination of the central values $\theta_1$, $\theta_2$ and  
$\phi_{12}$ the experimental data were integrated 
(cf.~Sec.~\ref{secIIB1}) within the limits of $\pm$~1$^{\circ}$ for 
the polar angles and of $\pm$~5$^{\circ}$ for the relative azimuthal 
angle.  The bin size along the kinematic curve $S$ was 4~MeV.  
Such limits allowed us to reach sufficient statistical accuracy while 
keeping the angle and energy integration effects to a minimum,
not affecting the comparison with the point-geometry theoretical 
predictions (see below).

\begin{figure*}
\includegraphics[width=168mm]{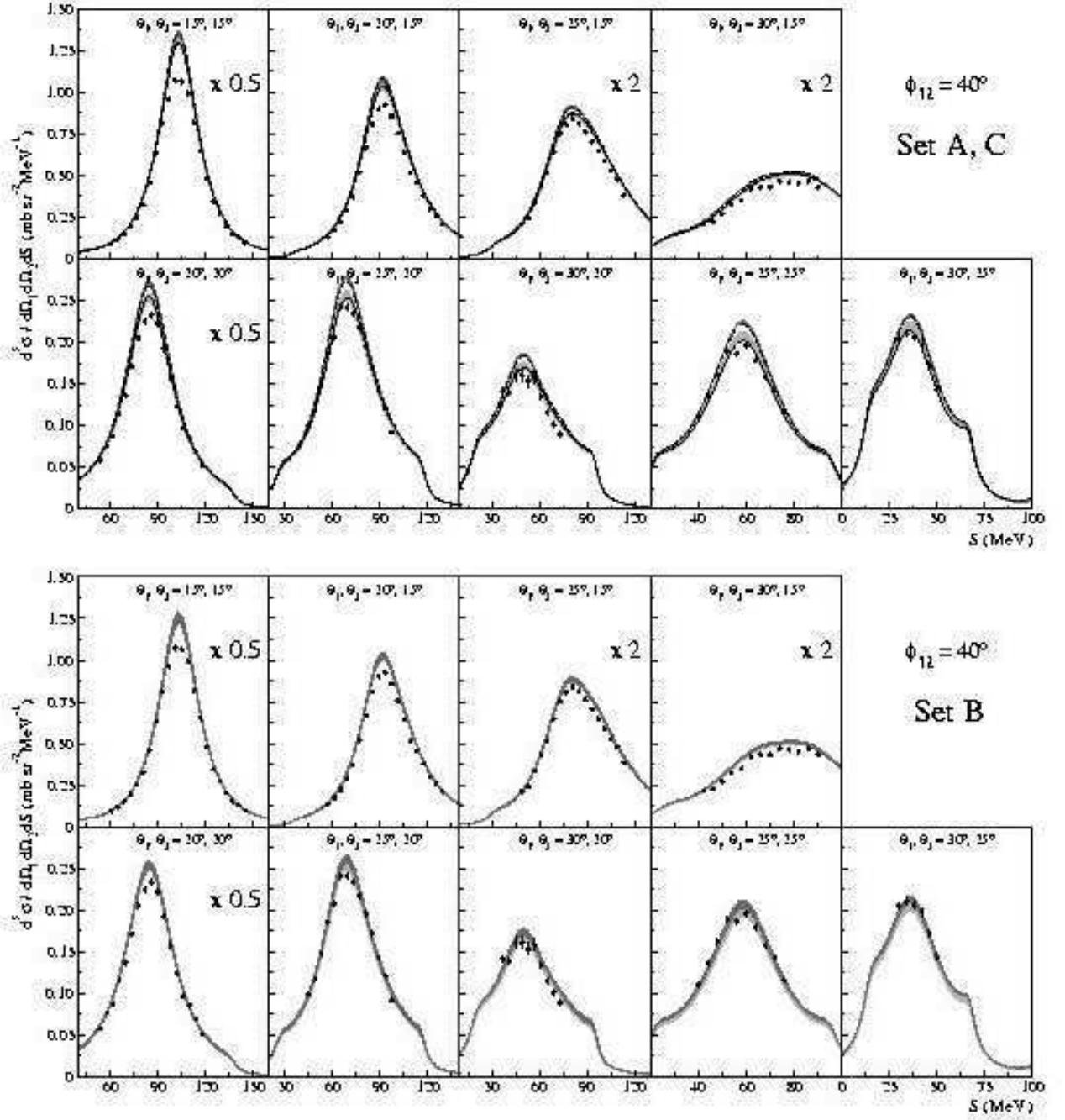}
\caption{\label{fig_res1}
Experimental breakup cross sections in 9 kinematical configurations for the relative azimuthal angle of the two protons 
$\phi_{12}$ = 40$^{\circ}$ and for various combinations of the proton 
polar angles, as indicated in the individual panels.  
The error bars represent statistical uncertainties only.  
In a few panels the results are scaled with the indicated scaling 
factors to fit the common vertical axis.  
{\bf Upper part}: data compared to predictions obtained with the 
realistic $NN$ potentials only (light-shaded bands), with calculations 
of the pairwise $NN$ forces combined with the TM99 3NF
(dark-shaded bands) and of AV18 + Urbana IX (dashed lines).  
The solid line represents the results obtained for calculations 
within the coupled-channel framework with the CD Bonn + $\Delta$ 
potential.
{\bf Lower part}: the same data confronted with the predictions 
obtained within the ChPT approach at NNLO.  The complete
calculations are represented by the dark-shaded bands, while the 
light-shaded ones demonstrate the results with the dynamics 
constrained to only $NN$ contributions.}
\end{figure*}

\begin{figure*}
\includegraphics[width=168mm]{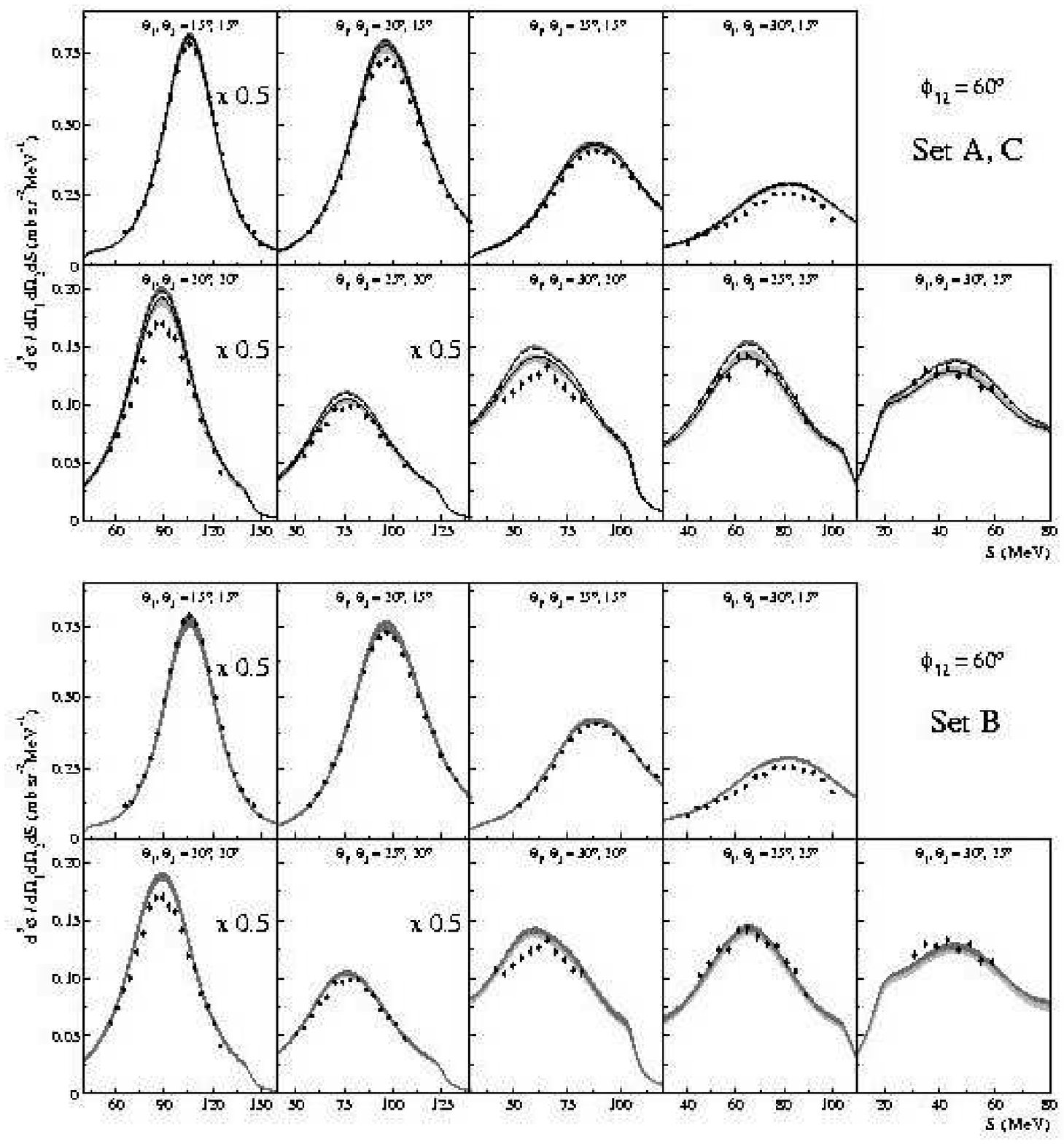}
\caption{\label{fig_res2}
The same as in Fig.~\ref{fig_res1} but for kinematic
configurations with the relative azimuthal angle of two protons
$\phi_{12}$ = 60$^{\circ}$.}
\end{figure*}

\begin{figure*}
\includegraphics[width=168mm]{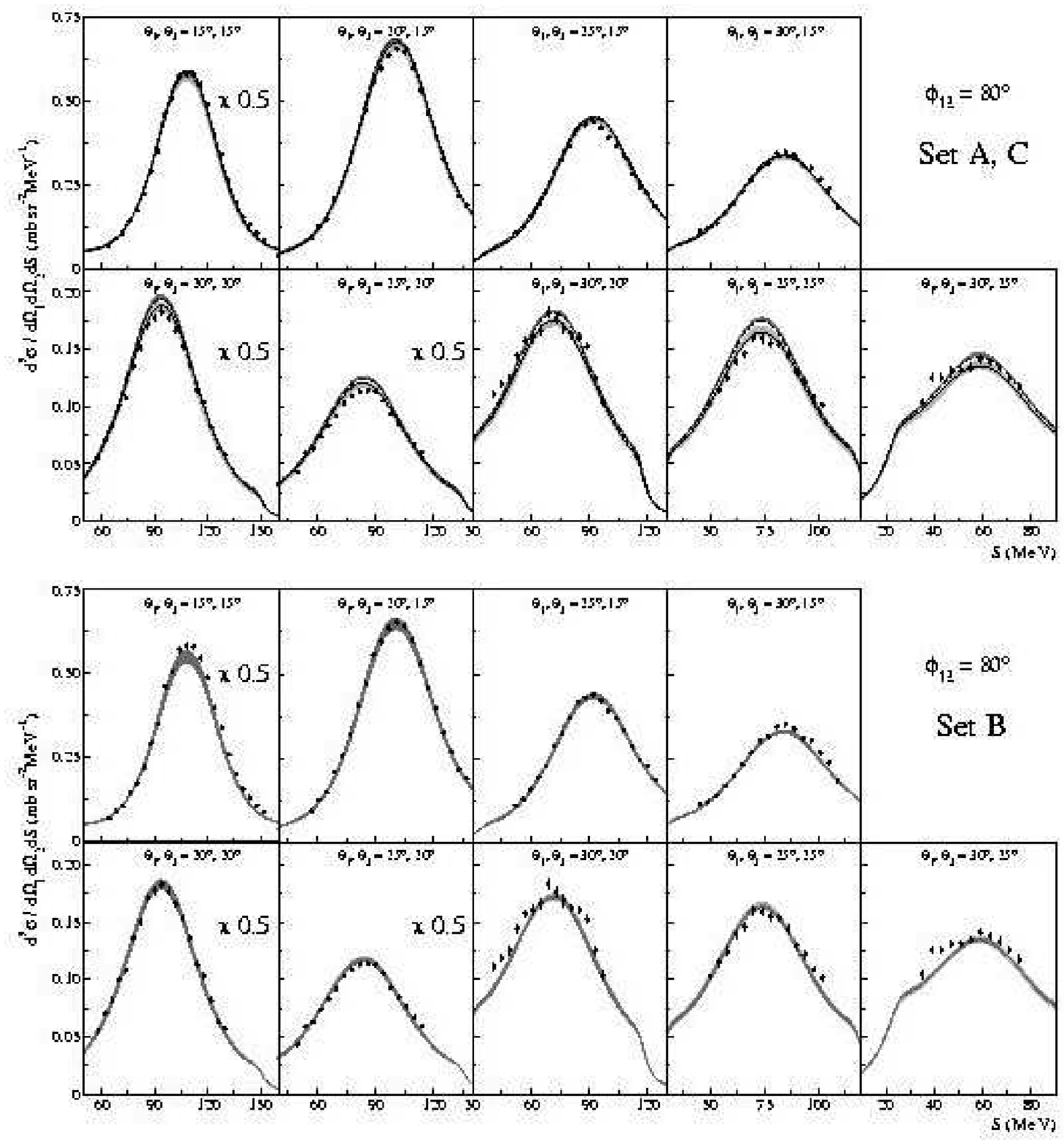}
\caption{\label{fig_res3}
The same as in Fig.~\ref{fig_res1} but for kinematic
configurations with the relative azimuthal angle of two protons
$\phi_{12}$ = 80$^{\circ}$.}
\end{figure*}

\begin{figure*}
\includegraphics[width=168mm]{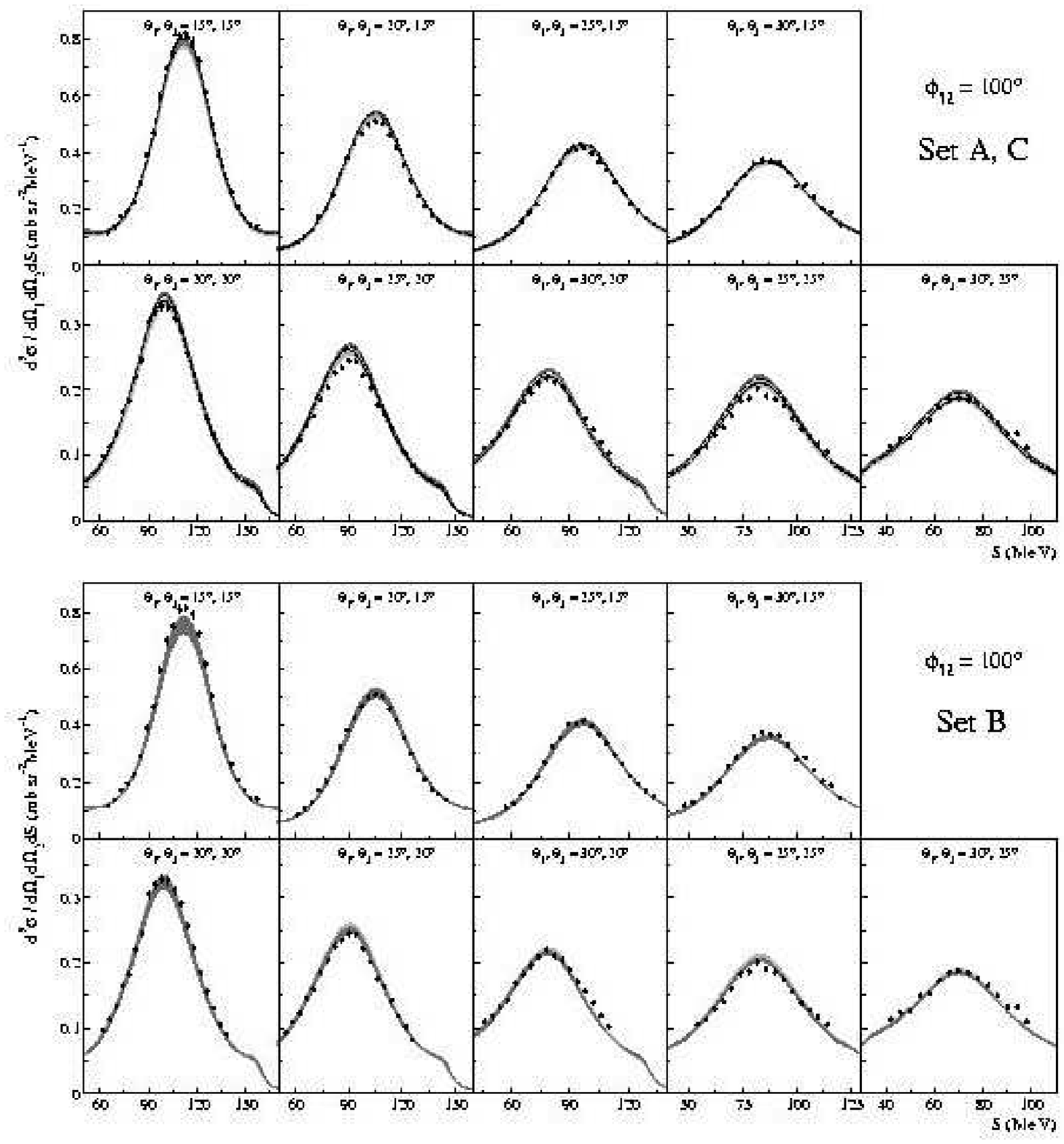}
\caption{\label{fig_res4}
The same as in Fig.~\ref{fig_res1} but for kinematic
configurations with the relative azimuthal angle of two protons
$\phi_{12}$ = 100$^{\circ}$.}
\end{figure*}

\begin{figure*}
\includegraphics[width=168mm]{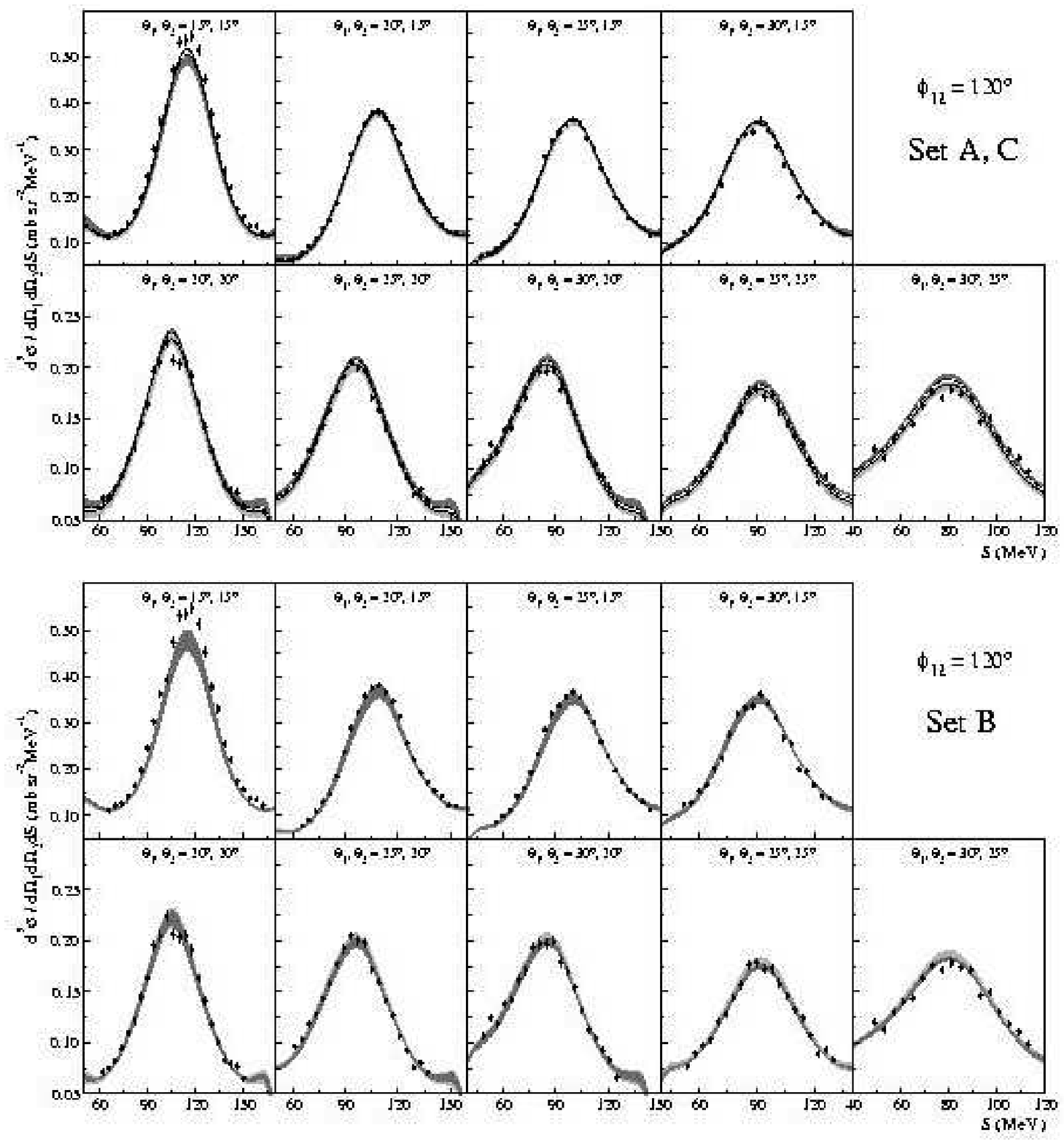}
\caption{\label{fig_res5}
The same as in Fig.~\ref{fig_res1} but for kinematic
configurations with the relative azimuthal angle of two protons
$\phi_{12}$ = 120$^{\circ}$.}
\end{figure*}

\begin{figure*}
\includegraphics[width=168mm]{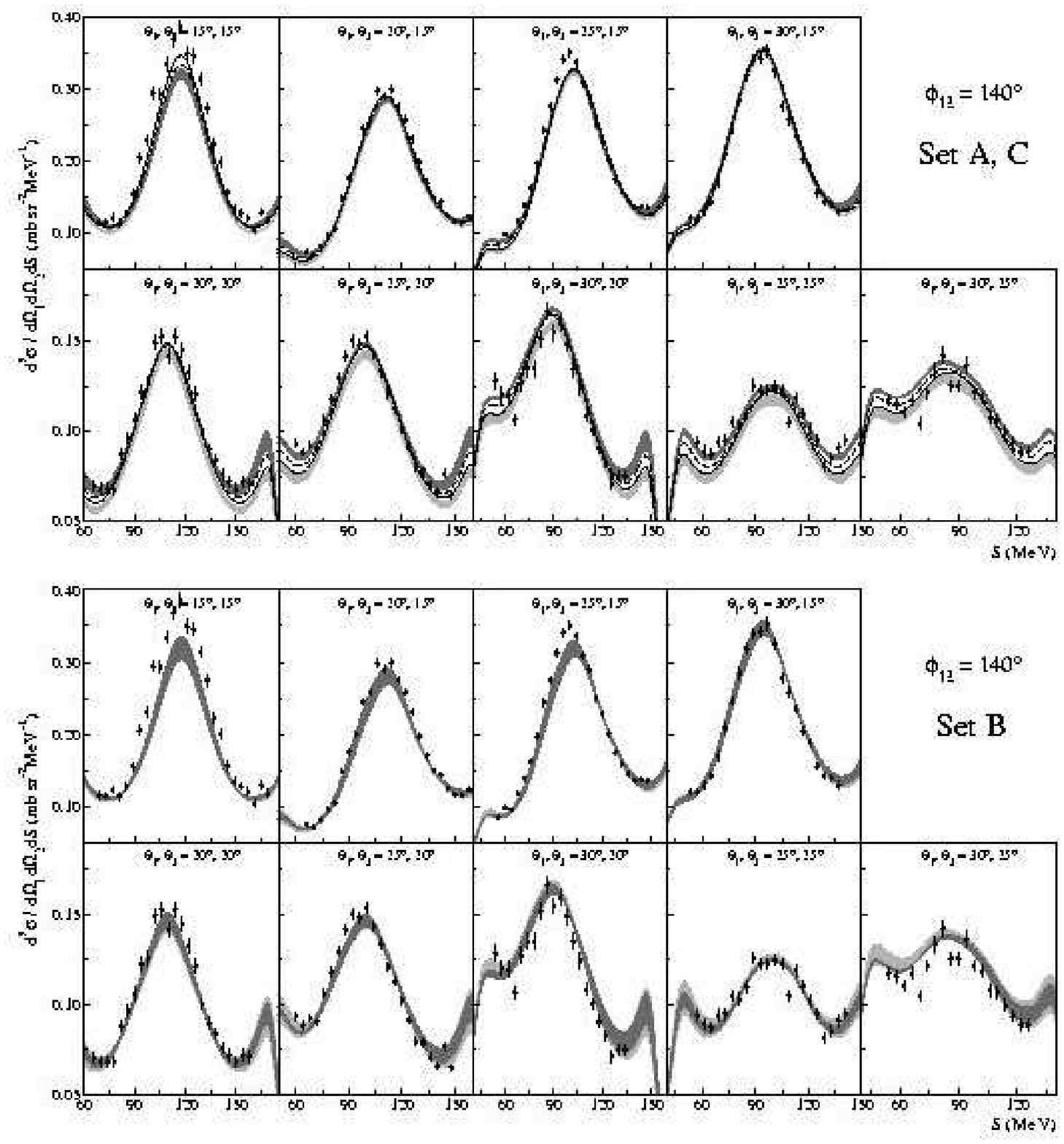}
\caption{\label{fig_res6}
The same as in Fig.~\ref{fig_res1} but for kinematic
configurations with the relative azimuthal angle of two protons
$\phi_{12}$ = 140$^{\circ}$.}
\end{figure*}

\begin{figure*}
\includegraphics[width=168mm]{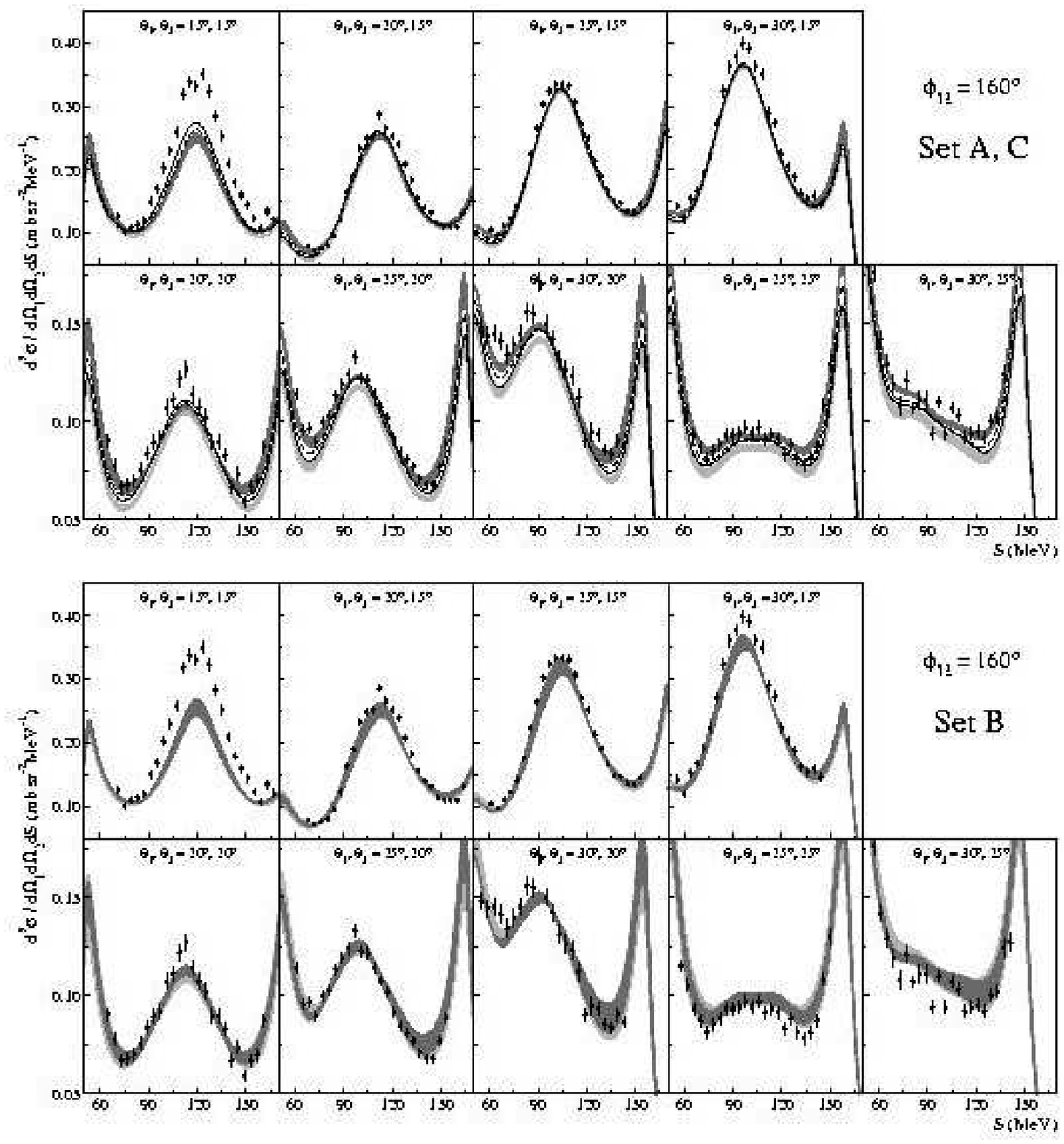}
\caption{\label{fig_res7}
The same as in Fig.~\ref{fig_res1} but for kinematic
configurations with the relative azimuthal angle of two protons
$\phi_{12}$ = 160$^{\circ}$.}
\end{figure*}

\begin{figure*}
\includegraphics[width=168mm]{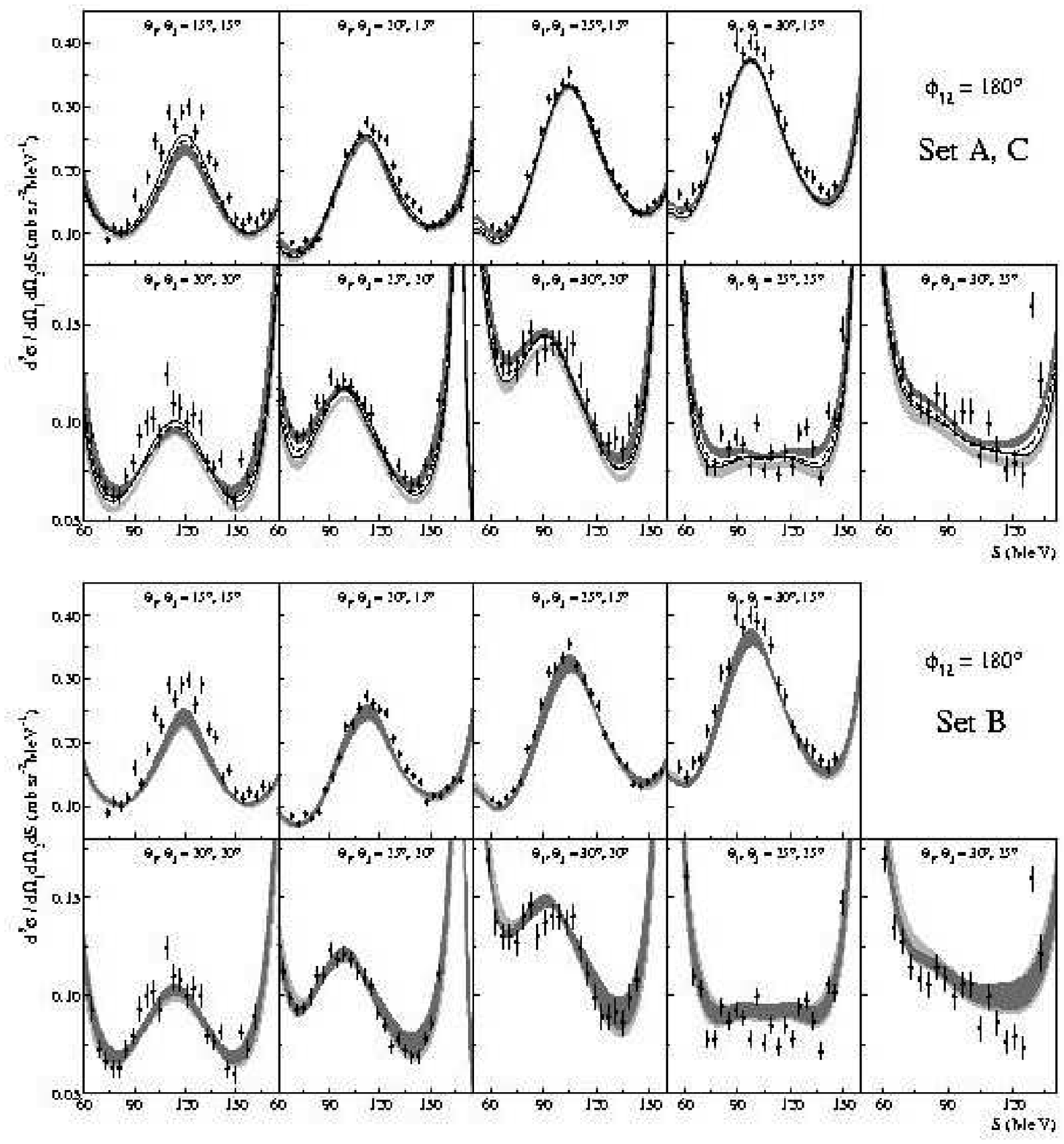}
\caption{\label{fig_res8}
The same as in Fig.~\ref{fig_res1} but for kinematic
configurations with the relative azimuthal angle of two protons
$\phi_{12}$ = 180$^{\circ}$.}
\end{figure*}

\subsection{\label{secIVA}Individual kinematical configurations}

A few geometries on the above defined grid were already presented in our 
previous report~\cite{Kis03a}.  However, due to improvements in the 
analysis procedure discussed in Sec.~\ref{secIIB}, the current
results are slightly more precise.  Therefore, and for
the sake of presenting a complete picture of data comparison with 
various theoretical approaches, we display in 
Figs.~\ref{fig_res1}-\ref{fig_res8} cross sections for all 
72 kinematically complete configurations, each figure showing
a collection of 9 geometries (different $\theta_1$, $\theta_2$ pairs)
for the same value of the relative polar angle $\phi_{12}$.  
The data are compared with three sets of theoretical calculations, 
introduced in Sec.~\ref{secIII}.
We refer to them by letters, corresponding to the numbering of the
respective subsection: the realistic potential approach with model 3NF's
(Sec.~\ref{secIIIA}) is called ``set A'', the ChPT predictions 
(Sec.~\ref{secIIIB}) are denoted by ``set B'' and the results for the 
coupled-channel potential with the explicit $\Delta$-isobar treatment
(Sec.~\ref{secIIIC}) are called ``set C''.  
Since the predictions of the three sets are often
close to each other, in order to clearly demonstrate all the details,
every figure is composed of two parts.  In the upper part the data 
are confronted with the results of calculations of set A and set C.
The light-shaded bands correspond to predictions obtained with only
pairwise $NN$ potentials (AV18, CD Bonn, Nijm I and II), the dark-shaded 
bands show the results when they are combined with the 2$\pi$-exchange 
TM99 3NF.  The dashed lines demonstrate the results of calculations  
with the AV18 potential combined with the Urbana IX 3NF.  The solid
lines show the predictions obtained with the use of the coupled-channel
potential CD Bonn + $\Delta$.  In the lower part the same data are
shown with the predictions obtained at NNLO of the ChPT approach.  
The bands show the ranges of the results computed using the  
different cut-offs, listed in Eq.~(\ref{eqChLmbd}); the light-shaded 
bands display the results when the calculations were restricted to 
include only the $NN$ force contributions, the dark-shaded bands 
represent the predictions for the full dynamics, with the 3NF graphs 
taken into account.
Following the arguments of our previous study~\cite{Kis03a}, we compare 
the experimental data averaged over finite phase-space intervals with 
the point-geometry theoretical predictions calculated at the central 
values of the ranges of the kinematical variables.  It has been checked that
for all the configurations considered here the averaging leads to a slight
enhancement of the theoretical cross-section values, not exceeding 
1.6\%, equivalent to some extra normalization factor.  Since the 
global conclusions are drawn mainly with eliminated influence of the 
data normalization (see further below), this simplification does not 
affect them.

Figs.~\ref{fig_res1}-\ref{fig_res8} are the basis for the quantitative 
comparisons of our experimental results with the predictions obtained in 
different approaches, as well as between the theoretical calculations
themselves.

There is a large number of configurations, concentrated mainly (but
not exclusively) in the central region of the investigated azimuthal 
angle $\phi_{12}$ range (Figs.~\ref{fig_res4}-\ref{fig_res5}, top 
panels in Figs.~\ref{fig_res3},\ref{fig_res6}-\ref{fig_res8}),
where predictions of all considered theoretical approaches are 
consistent with each other over the whole arclength range attainable in 
our experiment.  This is particularly true for geometries characterized 
by relatively large values of the cross section.
The bands representing ranges of cross section predicted by calculations
with different realistic potentials (set A) converge practically to a 
common line, identical with the predictions of the coupled-channel
potential (set C).  Similarly, the bands reflecting the computation
uncertainty of the ChPT approach (set B) are also very narrow.
Generally, in those configurations the theoretical predictions
follow very accurately the experimental distributions.  This confirms 
the high quality of the predictions provided by modern formalisms
and simultaneously reflects the precision and accuracy of the 
experiment.  However, since the predictions with and without 3NF's 
are identical, no details of the 3$N$ system dynamics can be gained 
from those data.  

There are, however, regions of the phase space, where the results of the 
calculations incorporating 3NF contributions differ substantially
from the ones using the $NN$ dynamics only. 
The most pronounced 3NF manifestations can be observed in the range  
$\phi_{12} > 120^{\circ}$, in the configurations characterized by 
relatively small cross sections (Figs.~\ref{fig_res6}-\ref{fig_res8}).
The induced changes concern the shape and/or the absolute magnitudes 
of the cross-section distributions.
High sensitivity of the predicted cross sections to the details of the 
interaction model applied in the calculations makes this region 
extremely useful for studying the 3$N$ system dynamics.
Regarding the results of the realistic potentials approaches (sets A 
and C), one observes that the inclusion of 3NF's usually increases 
the predicted cross-section values and that the largest effects are 
introduced by the TM99 3NF model, slightly smaller ones for the case of
Urbana IX 3NF and significantly smaller ones by the explicit 
$\Delta$-isobar excitation.  The comparison of the calculated cross
sections with the data leads to the important conclusion that the 
predictions of the realistic potentials describe the data much 
better when the contributions of the 3NF are taken into account.
The ChPT results do not reveal that clear signal of the 3NF effects.  
Inclusion of the 3$N$ interaction components, if affecting the predictions 
at all, results in a small change of the shape of the cross-section 
distribution along $S$ rather than in modification of its absolute 
magnitude.  In these geometries the bands representing the uncertainties 
of the ChPT predictions without and with the 3NF contributions are 
relatively wide, they essentially overlap one another, and generally 
their agreement with the data is satisfactory.  One can also observe 
that the calculated cross sections practically coincide with the 
results of the realistic potential approach containing the full 
dynamics, i.e.\ with the 3NF model included.

In geometries with small azimuthal angle, $\phi_{12} \leq 80^\circ$
(Figs.~\ref{fig_res1}-\ref{fig_res3}), one can also identify several 
cases where the contributions of the TM99 or Urbana IX 3NF's modify 
the predictions of the realistic potentials' approach at an appreciable 
level.  However, there are few reasons to consider the situation in this 
region as qualitatively different from that at large  $\phi_{12}$ values.  
The coupled-channel calculations with $\Delta$-isobar excitation 
included predict cross-section values consistent with those obtained 
for realistic $NN$ potentials without the 3NF contributions.  The set C 
predictions even tend to follow the lowest range of the set A band 
(see e.g.\ configuration $\theta_{1}$ = 25$^{\circ}$, 
$\theta_2$ = 25$^{\circ}$,  $\phi_{12}$ = 40$^{\circ}$ 
in Fig.~\ref{fig_res1}).  
Although the ChPT predictions without and with 3NF contributions still 
significantly overlap, ranges of both kind of predictions are relatively 
small.  As for set C, the set B predictions also agree rather well with the 
realistic potentials' results, which do not include 3$N$ interaction effects.
The comparison of theoretical results with the data shows noticeable
disagreements for several geometries.  Generally, all approaches, 
even if their results without and with 3NF contributions are almost
identical (see e.g.\ configuration $\theta_{1}$ = 30$^{\circ}$, 
$\theta_2$ = 15$^{\circ}$, $\phi_{12}$ = 40$^{\circ}$ in 
Fig.~\ref{fig_res1}), overestimate the data.  In geometries for
which the two kinds of predictions differ, this inconsistency
is worse for calculations with the 3NF contributions (TM99 or 
Urbana IX) taken into account, since adding this piece of dynamics
increases the predicted cross-section values.

The discrepancy mentioned above is the largest for the configuration
$\theta_{1}$ = 15$^{\circ}$, $\theta_2$ = 15$^{\circ}$,  
$\phi_{12}$ = 40$^{\circ}$ (first panel of Fig.~\ref{fig_res1}).
All theoretical approaches (although the effect is slightly less 
pronounced for the ChPT predictions) deviate from the data by as much 
as 20\%, i.e.~far beyond the experimental uncertainties.  Regarding all
configurations characterized by the smallest analyzed proton polar angles, 
$\theta_1 = \theta_2$ = 15$^\circ$, one finds that the disagreement
between the predictions and the experimental cross section changes
systematically, with a strong dependence on the relative azimuthal angle:
for the small $\phi_{12}$ values the data are overestimated, while
for the large $\phi_{12}$ they are strongly underestimated.  Only
for $\phi_{12}$ around 100$^{\circ}$ does the agreement become 
satisfactory.  To a much smaller extent this effect is visible also 
for geometries with $\theta_1 = \theta_2$ = 20$^\circ$ and perhaps 
$\theta_1$ = 20$^{\circ}$, $\theta_2$ = 15$^\circ$. 
It should be noted that for all configurations characterized by
a certain $\theta_1$, $\theta_2$ pair the same part of the detector 
is used to extract the data for any $\phi_{12}$ angle and, moreover, 
the efficiency corrections are tiny in comparison to the observed 
discrepancies, so it is impossible to attribute the inconsistencies 
to any experimental deficiency.  

\begin{figure}
\includegraphics[width=86mm]{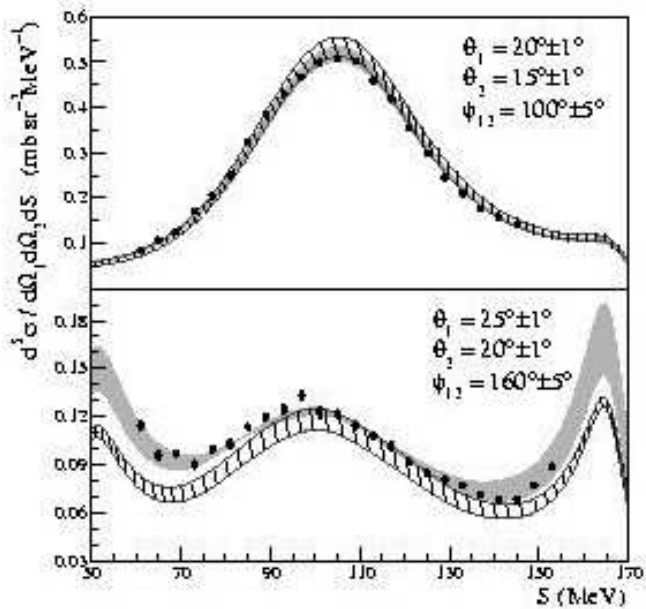}
\caption{\label{fig_n3n2}
Comparison of experimental breakup cross sections in two kinematic 
configurations (indicated in the panels)
with the predictions of ChPT performed at NNLO (dark-shaded bands) 
and at N$^{3}$LO (hatched bands).  Both calculations take 
into account only $NN$ force contributions.}
\end{figure}

The presented systematic study, covering a large fraction of the 
breakup phase space leads to a rather complicated picture.  Generally, 
for most of the studied geometries the description of data provided by 
all theoretical approaches is quite satisfactory.  There are specific 
regions, where the 3NF effects are pronounced and their importance is 
clearly confirmed by the measured cross sections. 
There are also final state geometries in which significant discrepancies
between the experiment and the theoretical predictions are observed. 
The pattern of disagreement changes as a function of the kinematical
variables.  These findings strongly support the statement that only 
precise measurements in large regions of the phase space can provide 
enough information to judge on the quality of the models dealing
with the description of the breakup observables.  Resolving the 
discrepancies is at present not possible; they might be a signal 
of some missing ingredients in the assumed dynamics of the 3$N$ 
system. 

Extending the investigation of the ChPT approach beyond the order 
discussed until now, we have compared our data with the predictions 
including only $NN$ contributions, obtained at the still higher, 
N$^{3}$LO order -- see~\cite{Epe05a} for details of the theory 
involved.
Since the absence of 3NF contributions makes these calculations by
virtue incomplete, we include the (incomplete) N$^{3}$LO predictions 
only in global $\chi^2$ tests (see below).  
In Fig.~\ref{fig_n3n2} we present sample 
comparison of predictions obtained at NNLO (dark-shaded) and N$^{3}$LO 
(hatched) for two geometries of the breakup reaction, both based
on the $NN$ potential only.  Comparing calculations based on 
incomplete dynamics can not be very conclusive, yet we can observe 
the differences between the obtained shapes, what may signal the 
importance of the higher order terms.  It is also expected that
the contributions of the 3NF at N$^{3}$LO might be larger
than at NNLO.  The quantitative comparison, however, must be
postponed until the full dynamics of the 3$N$ system is implemented
at that order.

\subsection{\label{secIVB}Global comparisons}

In order to perform a quantitative comparison of the whole bulk of our 
data with the theoretical predictions and to trace possible regularities 
in (dis-)agreement between data and theory, we continued the global tests
initiated previously~\cite{Kis03a}, calculating values of $\chi^2$ per 
degree of freedom between the data and individual sets of theoretical 
predictions.  We decided to concentrate on the option with a free 
normalization factor, putting more weight onto the shapes of the 
cross-section distributions as a function of $S$.
In this way the conclusions are not biased by the absolute normalization
uncertainties.
In a later part of this Section we describe a complementary piece of 
investigation, presenting a comparison of data integrated over the 
arclength variable $S$ with the analogously treated theoretical 
predictions.  There the experimental normalization is fully taken 
into account, while the dependence on $S$ (i.e.\ shape of the 
distribution) is to a large extent neglected.  
This approach is an example of studying the breakup phase space by 
inspecting its projections onto selected sub-spaces of lower dimensions.

Focusing our comparison on how the shapes of the cross-section 
distributions are reproduced by different theoretical approaches, 
we have calculated the values of $\chi^2$ per d.o.f.\ for all 72 
configurations together (a total of nearly 1200 cross-section data points) 
with respect to all sets of theoretical predictions.  
The emphasis on the shapes of the distributions is motivated by the
observation (cf.~Sec.~\ref{secIVA}) that the action of 3NF contributions
is usually equivalent to a small increase of the cross-section values 
in the whole range of $S$.  Therefore, if the experimentally determined 
absolute normalization factor would be e.g.\ slightly too large, the data 
would be artificially shifted towards the predictions including full 
dynamics, leading to erroneous conclusions.  This method also allows us 
to eliminate the small influence of averaging, inherently present 
in the data and omitted in the theoretical predictions (averaging does 
not affect the shapes of the cross-section distributions presented here).
To eliminate the influence of the absolute experimental normalization, 
the data were renormalized in each configuration by a constant
factor (limited to the range 0.9 to 1.1), to best fit the particular
theoretical distribution.  In this way the quality with which a given
set of theoretical predictions reproduces all the data is quantified
by a single number.  In particular, for each combination of forces
we can compare two $\chi^2$ values: $\chi^{2}_{2N}$ obtained for
predictions based on pairwise $NN$ interaction only and 
$\chi^{2}_{2N+3N}$ for the calculations including 3NF contributions.
It should be noted that in the $\chi^2$ analysis only statistical 
uncertainties were taken into account, therefore values exceeding  
1 can be expected.  Investigating influences of the 3NF 
effects we concentrate rather on the {\em relative}\/ change from 
$\chi^2_{2N}$ to $\chi^{2}_{2N+3N}$ and not on the absolute $\chi^2$ 
value.  The same argument holds when comparing predictions of
different forces with respect to the quality with which they
describe the experimental data.

\begin{table}
\caption{\label{tab_chi2}
Agreement between the experimental cross sections and the theoretical 
predictions obtained in different approaches, quantified in terms of 
$\chi^2$ per degree of freedom.  Major focus in comparing the data
with theory is put on the shapes of the experimental distributions as
explained in the text.  The quality of the predictions based on only
pairwise $NN$ interactions ($\chi^{2}_{2N}$) is compared with 
$\chi^{2}_{2N+3N}$ values obtained for calculations including genuine 
3NF effects, typical for the particular approach.  Details on the
kinds of forces used in obtaining $\chi^2$ values at every row are
given in the text.}
\begin{ruledtabular}
\begin{tabular}{llcc}
$NN$ force & 3NF model & ~~~~~$\chi^{2}_{2N}$~~~~~ & ~~~$\chi^{2}_{2N+3N}$~~~\\
\hline
\multicolumn{4}{c}{Set A} \\
AV18           & TM99      &  4.48        &  3.80 \\
~~~~           & Urbana IX &  ~~~~        &  3.67 \\
CD Bonn        & TM99      &  4.04        &  3.80 \\
Nijm I         & TM99      &  4.38        &  4.38 \\
Nijm II        & TM99      &  4.53        &  4.05 \\
Mean realistic & TM99      &  4.43        &  4.07 \\
~~~~           & ~~~~      &  (4.04-4.53) &  (3.80-4.38) \\
\multicolumn{4}{c}{Set B} \\
\multicolumn{2}{l}{ChPT at NNLO} &  3.67      &  3.96 \\
~~~~           & ~~~~      &  (3.36-6.59) &  (3.35-6.64) \\
\multicolumn{2}{l}{ChPT at N$^{3}$LO}   &  5.29      &  -- -- \\
\multicolumn{4}{c}{Set C} \\
Coupled-channel & $\Delta$-excitation &  3.83  &  3.63 \\
\end{tabular}
\end{ruledtabular}
\end{table}

The results of the $\chi^2$ analysis for all considered theoretical
approaches are shown in Table~\ref{tab_chi2}.  First, the two kinds
(without and with 3NF contributions included in the theory) of $\chi^2$ 
values are shown for 4 realistic $NN$ potentials and their combination
with the TM99 3NF.  The second row for the AV18 potential gives 
$\chi^{2}_{2N+3N}$ for this force combined with the Urbana IX 3NF.
Excluding this last combination, we define a ``mean realistic'' 
prediction as a set of cross-section values given at each point 
($\theta_1$,$\theta_2$,$\phi_{12}$,$S$) as a mean between the minimum
and maximum cross section predicted by the 4 realistic forces (or their
combination with TM99 3NF) at this kinematical point.  The $\chi^2$ 
values with respect to this mean realistic prediction are also shown
in Table~\ref{tab_chi2}, with the ranges, equal to the corresponding 
extreme $\chi^2$ values, repeated in the next row.
These values are to be compared with the ones obtained for the
ChPT calculations.  In this case only the $\chi^2$ for the ``mean'' set 
is quoted.  It is obtained analogously as in the realistic potentials
case, as the central value between extremes predicted with 5 combinations
of cut-offs.  Also the ranges of $\chi^2$ values for predictions at 
NNLO are shown for comparison.  In the case of N$^{3}$LO calculations 
we quote only the central $\chi^2$, reminding the reader that the bands 
based on the $NN$ forces only tend to be wider at this computational order 
and therefore also the accuracy of the predictions would behave accordingly.  
The last row of Table~\ref{tab_chi2} presents the results of $\chi^2$ 
analysis performed with respect to the coupled-channel potential.  
A small difference in $\chi^2_{2N}$ values for CD Bonn between set A 
and set C calculations is due to the different treatment of the charge 
dependence.  The $\chi^{2}_{2N+3N}$ denotes here the value obtained for 
the calculations including all $\Delta$-isobar excitation effects.

Comparing the numbers presented in Table~\ref{tab_chi2} one observes
that combining any of the realistic potentials with a 3NF model
improves the description of our data, decreasing the $\chi^2$ by 
about 10\%.  The effect varies for different $NN$ potentials of the set A
calculations so that the ranges of $\chi^{2}_{2N}$ and $\chi^{2}_{2N+3N}$
values overlap.  Nevertheless, a systematic shift of the predictions
towards the data is visible for calculations including the full dynamics.
Predictions obtained within the ChPT framework do not allow for such
a conclusion.  Although the quality of the data description for the set B 
calculations is very similar to that of the realistic potentials approach, 
the ranges of $\chi^2$ values obtained with and without 3NF 
contributions are very wide and overlap completely.  The central-value 
predictions reveal even a slight worsening of the data description
induced by including the 3NF effects.  However, this observation
is rather far fetched in view of large theoretical uncertainties
and, moreover, taking the center of the relatively wide bands
is not relevant for tracing details of the shapes of the distributions.
The predictions of $NN$ force alone at N$^{3}$LO show a much poorer
agreement with the data than those of NNLO.  Since the total 3NF
contributions up to that order are expected to be larger than at NNLO,
this effect might be easily compensated for by the complete dynamics
included in the formalism.
The set C calculations, with the coupled-channel potential and explicit
$\Delta$-isobar degrees of freedom, lead in general to the smallest 
values of $\chi^2$.  The effects of the $\Delta$ excitation are rather 
modest, but with their inclusion the predictions are moved by a few 
percent closer to the data.  This conclusion is, however, biased by
the absence of any estimate of the theoretical uncertainties.

As mentioned in Sec.~\ref{secIVA}, the largest disagreements between
the data and the theoretical predictions are observed for configurations
with the smallest polar angles, $\theta_1 = \theta_2$ = 15$^{\circ}$.
In order to eliminate their possible dominant impact on the $\chi^2$
analysis, we have recalculated all the values of Table~\ref{tab_chi2}
excluding this piece of data (8 configuration out of 72).  It has been
found that the $\chi^{2}_{2N}$ and $\chi^{2}_{2N+3N}$ values obtained in 
that way were decreased only by about 5\%, but the overall picture was
preserved and thus all the above conclusions are valid also for 
that limited data sample.  Below we present less global comparisons
based on this subset of the data.

\begin{figure}
\includegraphics[width=86mm]{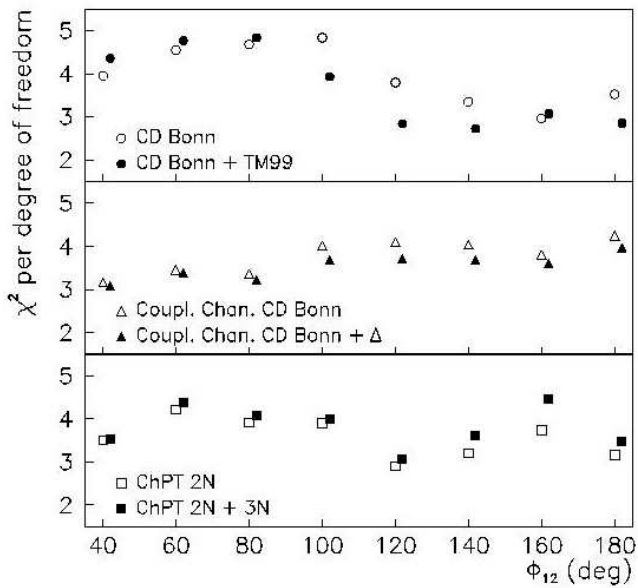}
\caption{\label{fig_chph_frenor}
$\chi^2$ per degree of freedom calculated for groups of kinematical 
configurations with the same value of $\phi_{12}$.  The results 
of data comparison with calculations of sets A, C and B are shown
with dots, triangles and squares, respectively, in the separate panels.  
Empty and full symbols correspond to predictions based on $NN$ forces 
only and with the 3NF contributions included.  For clarity the 
results for the full dynamics are artificially shifted along 
the $\phi_{12}$ axis by 2$^{\circ}$.}
\end{figure}

To search for possible regularities in changes of the quality of the data
description by the models, a less global treatment is obviously needed.
Firstly, we studied the consistency between the data and theoretical 
predictions in various regions of the phase space, inspecting 
the dependence of $\chi^2$ on the relative azimuthal angle $\phi_{12}$ 
of the two protons.  Values of $\chi^2$ have been 
calculated as in the global comparison case, but for the groups 
of configurations characterized by the same $\phi_{12}$ value, 
i.e.~separately for each group presented in 
Figs.~\ref{fig_res1}-\ref{fig_res8}.  The results are shown in 
Fig.~\ref{fig_chph_frenor} by three different symbols representing the 
three calculation sets and the $\chi^{2}_{2N}$ and $\chi^{2}_{2N+3N}$
shown by open and full symbols, respectively.  To simplify the picture 
the realistic potentials are represented by the CD Bonn force (with 
and without TM99 3NF) and the ChPT approach results are shown for the 
mean set.
One observes that for $\phi_{12}<90^{\circ}$ there is practically no 
effect of including the TM99 3NF into the calculations.  For larger 
relative azimuthal angles the description of data is significantly
improved by employing the full dynamics.  The coupled-channel
calculations predict much smaller effects due to $\Delta$ excitation,
what is the result of a compensation mechanism, cf.~Sec.~\ref{secIIIC}. 
Moreover, the quality of description depends only very weakly on  
$\phi_{12}$.  The ChPT predictions do not show any large effects
of 3NF contributions either.  Only for the largest azimuthal angles,
$\phi_{12}>120^{\circ}$, in contrast to the CD Bonn results, 
the full dynamics reproduces the data worse than calculations with 
$NN$ interaction terms only.

\begin{figure}
\includegraphics[width=86mm]{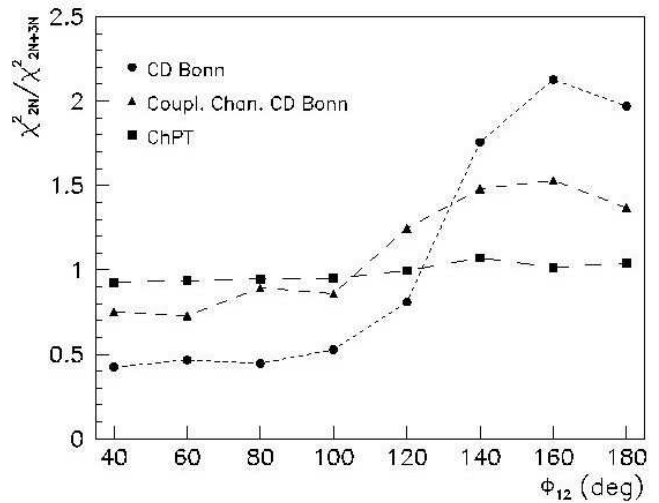}
\caption{\label{fig_chph_absnor}
Ratios of $\chi^2$ values for calculations without and with 3NF
contributions.  The cross-section data with the experimental
absolute normalization were used in computing the 
$\chi^2_{2N}$ and $\chi^2_{2N+3N}$ values for groups of kinematical 
configurations with the same $\phi_{12}$ angle.  The results 
of data comparison with calculations of sets A, C and B are shown
by dots, triangles and squares, respectively.  The lines are only 
to guide the eye.}
\end{figure}

Pursuing the study of the $\chi^2$ dependence on the relative azimuthal
angle, we have also checked the changes of quality of data description
by calculations with and without 3NF contributions for the cross
sections with absolute experimental normalization applied to the data. 
We present the results as the ratio of $\chi^2_{2N}$ to $\chi^2_{2N+3N}$ 
in order to magnify the influence of the $3NF$ effects.
In Fig.~\ref{fig_chph_absnor} the ratios for the same theoretical 
sets as for the free normalization case are shown with the
same symbols.  One finds again that the consistency between the 
predictions of the CD Bonn potential and the data is improved by 
adding TM99 3NF in configurations with relatively large $\phi_{12}$
angles (ratio above 1).  On the other hand, for $\phi_{12}<90^{\circ}$ 
including the 3NF into the calculations moves the results away from 
the data.  Astonishingly, the magnitude of the relative change is
almost the same in both directions, described by a factor of about 2.  
For the ChPT calculations no effect is present, the ratio stays close to 
1 for all values of $\phi_{12}$.  The behavior revealed by the set A 
predictions is qualitatively confirmed by the coupled-channel 
calculations, however the amplitude of the changes induced by including 
the $\Delta$-excitation contributions is smaller.

A great advantage of an experiment with the position sensitive detector
covering a significant part of the phase space is the opportunity to
study dependences of the observables (here the differential cross 
section) on all independent kinematic variables.  However, inspecting
the results in many-dimensional space is difficult and the comparisons
with the theoretical predictions might miss the regularities.
One possible solution to reduce the complexity of the problem and still 
be able to use all the data is to select a small number (e.g.\ 1 or 2) 
of variables and to  
integrate the observable over the others.  Integration of the experimental 
data is usually quite straightforward -- it is accomplished by summing up 
of events which fulfill the required conditions.  But these experimental 
conditions (acceptances, thresholds, granularity, etc.) impose limitations, 
which make the procedure of integration for the theoretical predictions
very complicated and the comparisons might be jeopardized by introducing
uncontrollable systematic errors.
A possible method to resolve the problem of comparing the integrated 
experimental and theoretical observables has been suggested 
in~\cite{Kur04a}.  It allows to effectively integrate the calculated
observables over all but one variables, with all experimental
constraints taken into account.  In the case of cross sections, however, 
such an approach leads to numerical values which are hard to interpret 
physically.  Influences of the physical changes (due to reaction 
dynamics) of the observable are merged with the acceptance functions
and the comparisons are meaningful only for ``integrated physical
values''.  Therefore, in our first attempts to investigate regularities
in the breakup phase space we rather employ a simpler method,
deconvoluting the acceptances from the experimental results
and comparing the integrated {\em cross sections}\/ with the accordingly 
summed theoretical predictions.  In this way we end up with
``objective'' cross-section values, with direct physical interpretation.

\begin{figure}
\includegraphics[width=78mm]{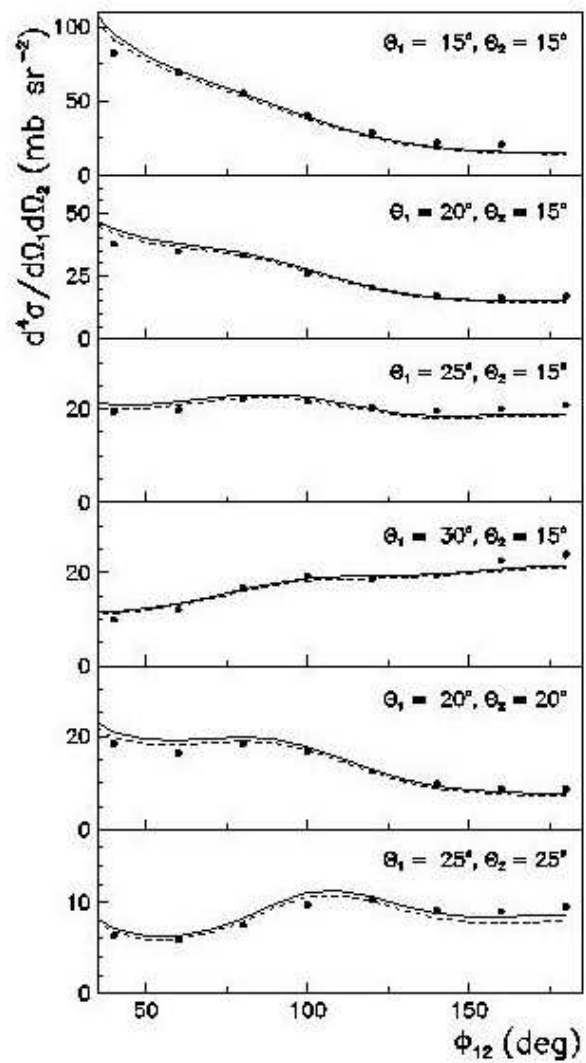}
\caption{\label{fig_csint}
Differential, integrated over $S$, cross-section values, presented as 
functions of the relative azimuthal angle $\phi_{12}$, for several 
pairs of the proton polar angles $\theta_1$ and $\theta_2$ (indicated
in the panels).  The integration limits are defined by setting the proton 
energy threshold at 25 MeV.  The data points are compared with the
results of calculations with the CD Bonn potential (dashed lines) and
with the CD Bonn + TM99 3NF combination of forces (solid lines).}
\end{figure}

The cross-section results for the individual configurations shown in 
Figs.~\ref{fig_res1}-\ref{fig_res8} suggest a possible 
correlation between the polar and azimuthal angles with respect to
the quality of the agreement between the data and predictions. 
Therefore, we studied the cross-section dependences on the proton 
emission angles with the experimental data integrated only over $S$.
The energy threshold of the detection system introduces
an inherent influence of the instrumental acceptance onto the result, 
but it is easy to reproduce without any detailed knowledge of other
features of the apparatus.  To guarantee an exact equivalence of the
low-energy cut-off condition for the experimental and the predicted 
results, a threshold of 25~MeV (higher than the hardware level) was 
applied for both proton energies and only the range of $S$ limited 
by this requirement was included in the integration.   
The results for 6 pairs of the proton polar angles as functions
of the relative azimuthal angle are shown in Fig.~\ref{fig_csint}.
The integrated experimental cross sections are compared to the
correspondingly integrated theoretical predictions
based on the CD Bonn potential only (dashed lines) and with the
TM99 3NF included in the calculations (solid lines).  It has been
checked that the results for other realistic forces are almost
indistinguishable from the ones presented in Fig.~\ref{fig_csint}.
The tendency, already visible in the cross-section plots for
individual configurations, can be better traced here. For $\phi_{12}$
below $90^\circ$ the theoretical predictions overestimate the data 
and the discrepancy rises with decreasing $\phi_{12}$ values. 
In the central region of the analyzed $\phi_{12}$ range the
agreement between the data and the theoretical curves is the best. 
With further increase of the $\phi_{12}$ angle the theoretical 
predictions start to underestimate the data.  This discrepancy
is, however, reduced (in various fractions) by including the 3NF 
into the calculations.  On the contrary, for the small $\phi_{12}$ 
angles the effects of 3NF inclusion increase the discrepancies 
between the predictions and the data.  It can be finally stated 
that in all cases studied in Fig.~\ref{fig_csint} the slope of the
data, though qualitatively reproduced, is not exactly matched by
the theoretical cross-section values and that overall rather
small 3NF contributions do not change the global picture.  
Again, this has to be attributed to some still unresolved
deficiencies of the models of the 3$N$ system dynamics.

Recapitulating all the results, from both the individual configurations of 
Figs.~\ref{fig_res1}-\ref{fig_res8} and the global tests from this 
section, we can conclude, that the present day models of the 3$N$
system dynamics reproduce the majority of the data with satisfactory
precision.  In many cases in which the predicted effects due to 
3NF's are non-negligible, their inclusion tends to improve the agreement
with the data.  However, thanks to the applied experimental technique
of covering a significant fraction of the breakup phase space with a
highly symmetric detection system, it has been shown that there are
also systematic regularities in the discrepancies between the measured 
cross sections and the predictions of all the theoretical approaches.
Since the systematic factors of our measurement are common to all
configurations, the established trends cannot be attributed to 
systematic experimental uncertainties and therefore hint at
missing ingredients of the nuclear Hamiltonian models.
It should also be stressed that additional complete and precise data sets, 
at other energies and in even larger phase space regions, are needed 
to study details of the interactions in the few-nucleon system.

\section{\label{secV}Summary and conclusions}

A measurement of the deuteron-proton $^1$H(d,pp)n breakup cross sections 
using a 130 MeV deuteron beam was performed for a large part of the 
available phase space.  In this paper high precision, five-fold differential 
cross-section data for 72 kinematically complete configurations (total
of nearly 1200 cross-section data points) at different angular combinations 
of the two outgoing protons are presented.  We discuss first examples of a 
global analysis of the data, which is trying to establish possible 
regularities of the (dis-)agreement between the experimental data and 
different theoretical approaches.

We compare the measured cross sections to theoretical predictions treating
the full dynamics of the 3$N$ system in three different ways:  employing
the realistic $NN$ potentials AV18, CD Bonn, Nijm I and Nijm II and 
including the 3NF effects by combining them with the TM99 3NF model 
(for AV18 also Urbana IX 3NF), obtaining the nuclear effective potential
in the ChPT approach with the calculations performed at NNLO order with
$NN$ and 3$N$ contributions (mentioning also the pure $NN$ results
obtained at N$^3$LO), and using the coupled-channel technique of explicit 
inclusion of a single $\Delta$-isobar degrees of freedom, resulting 
in a modified form of the realistic CD Bonn force and its combination 
with all single $\Delta$ excitation effects in the three-baryon system.  
The three approaches match equally well the properties of the $NN$ 
system.  When only the 2$N$ dynamical sector is used, their predictions
for the breakup cross sections are essentially equivalent.
    
The magnitude of the predicted 3NF effects depends on the approach.
In the case of the coupled-channel potential the influences of the
$\Delta$-isobar excitation are generally rather small.  This is due
to a competition of effects induced by two mechanisms, the two-baryon 
dispersion and the effective 3NF.  ChPT predictions, considered in 
terms of ranges of the cross-section values computed with different 
cut-off parameters, reveal also rather weak contributions of the 3NF 
effects.  They are usually smaller than the residual dependence on the
cut-offs.  Also the calculations at N$^3$LO (presently available with $NN$ 
contributions only) are characterized by a broad range for the predicted 
cross-section values.  The deviations of NNLO and N$^3$LO predictions
depend on the final state geometry.  
The largest sensitivity to the 3NF is found in the realistic potentials 
approach.  But even for the realistic forces there are several final state 
geometries in which the 3NF effects are practically negligible.  
Generally, in such cases the cross-section data are in good agreement 
with the theoretical predictions.  However, in many analyzed 
configurations the effects of including the 3NF are not negligible.  
Taking into account the 3NF contributions in the calculations leads  
to an increase of the cross-section values.  This effect is slightly 
less pronounced for combining the Urbana IX 3NF with the AV18 $NN$ 
potential than for the TM99 force combined with any of the four
considered $NN$ potentials.  

A global analysis, focused on the shapes of the cross-section distributions
as functions of the arclength variable $S$, shows that the agreement 
between the experimental data and the theoretical predictions improves
when the 3NF contributions are taken into account.  This conclusion is 
valid for all combinations of realistic $NN$ potentials with the model 
3NF's.  While for the ChPT predictions no conclusions can be drawn due 
to essentially overlapping ranges of predictions without and with 3NF
included, the coupled-channel calculations also reveal a slight 
improvement in the description of the data when the single $\Delta$-isobar 
excitation effects are incorporated.

There is, however, a number of configurations in which the cross-section
data are not correctly reproduced by any calculation.  
The effect depends on the relative azimuthal angle $\phi_{12}$ of the
two protons: for small values the data are overestimated by the 
predictions, the agreement becomes good in the central range of the 
analyzed $\phi_{12}$ and at the largest angles the discrepancy is reversed.
Thanks to the highly symmetric form of our detection system, which 
allows to reduce and to carefully control and test for systematic 
uncertainties, we basically exclude the possibility of attributing 
this inconsistency to any experimental deficiency.

The regularities of disagreements have been studied in more detail 
using a global analysis, in which we have concentrated on both,
the shapes of the distributions as well as on the absolute values 
of the cross sections.  It has been established that for configurations 
with large values of the $\phi_{12}$ angle the agreement between the 
experimental data and the theoretical predictions is improved 
when the 3NF contributions are taken into account.  On the contrary, for
$\phi_{12} < 100^{\circ}$ the 3NF effects move the
predictions away from the data.  This conclusion is valid for all 
combinations of realistic $NN$ potentials with the model 3NF's.
It also holds for the predictions obtained in the coupled-channel
approach, but with reduced size of the effects induced by the
$\Delta$-excitation.  The ChPT calculations predict essentially no 
sensitivity to the 3NF influences along $\phi_{12}$.

We have confirmed that sizable influences of 3NF's are visible in the 
breakup cross sections at the energy of our measurement.  
Comparison of the agreement between the experimental data and the 
predicted cross-section values is presented by first examples of 
analyzing a multi-dimensional breakup data set by inspecting its 
projections onto selected sub-spaces of lower dimensions.  
Since the advantages of the experimental method reduce strongly 
the impact of systematic errors, the established discrepancies 
might be considered as signal of some missing ingredients in the 
assumed dynamics of the 3$N$ system.  Determining regularities in
the disagreements might eventually help to identify shortcomings 
of the nuclear force models.  However, it cannot be ruled out that the
discrepancies result from Coulomb effects, which are ignored in all 
theoretical approaches presently used.  In view of the recent advances 
in including long range forces into the Faddeev formalism this
persisting question might soon be quantitatively addressed.

Our present study clearly shows the importance of complete, precise 
experiments, determining various observables of the breakup reaction.
Implementing symmetric detection systems covering large fractions
of the phase space allows to attain rich data sets, for which systematic
uncertainties are strongly suppressed and well controlled.  Results of 
such experiments are the basis for systematic comparisons with different 
theoretical approaches.  They provide stringent criteria for verification
of the models of the nuclear Hamiltonian, assumed in formulating the
three-nucleon scattering problem.
Further improvements of the theoretical models, which are also a basis 
for exact calculations in many-body systems, require still more 
experimental input.
We are going to supplement the cross-section results with polarization 
observables, with which we will be able to provide more detailed 
information, and hopefully pin down the discrepancies found here.
More data sets are needed, acquired at energies 
lower and higher than in our study, with proton and deuteron beams, 
as well as covering still larger fractions of the phase space.

\begin{acknowledgments}
This research was supported by the Polish Committee for Scientific Research 
under Grants No.\ 2P03B02818 and 2P03B00825.  
It is a part of the EU Integrated Infrastructure Initiative 
Hadron Physics Project under contract number RII3-CT-2004-506078. Work 
supported in part by DFG (SFB/TR 16, ``Subnuclear Structure of Matter'').
It has also been supported by the U.S. Department of Energy Contract 
No.\ DE-AC05-84ER40150 under which the Southeastern Universities Research 
Association (SURA) operates the Thomas Jefferson Accelerator Facility. 
The numerical calculations have been performed on the Cray SV1 and IBM 
Reggatta p690+ of the NIC in J\"ulich, Germany. 
The authors would like to express their appreciation for the tireless
efforts of the AGOR and the polarized source groups at KVI.  Polish
members of the experimental crew sincerely acknowledge support and  
hospitality of the KVI during the data taking periods.
\end{acknowledgments}

\end{document}